\documentclass[useAMS,usenatbib]{mn2e}
\usepackage{bethmacros}
\usepackage{epsfig}
\usepackage{epstopdf}
\DeclareGraphicsExtensions{.pdf, .jpg}

\title[SPH Star Formation and Feedback--I. Isolated Galaxies]{Star Formation and Feedback in Smoothed Particle Hydrodynamic Simulations--I. Isolated Galaxies} 
\author[Stinson et al.]{
{Greg Stinson$^{1}$\thanks{E-mail: stinson@astro.washington.edu}, Anil Seth$^{1}$, Neal Katz$^{2}$, James Wadsley$^{3}$, Fabio Governato$^{1,4}$, Tom Quinn$^{1}$
}
\vspace*{6pt}\\
$^{1}$Department of Astronomy, University of Washington, Box 351580, Seattle, WA 98195, USA\\
$^{2}$Department of Astronomy, University of Massachusetts, Amherst, 01003, USA\\
$^3$Department of Physics and Astronomy, McMaster University, Hamilton, Ontario L88 4M1, Canada\\
$^4$INAF, Osservatorio Astronomico di Brera, via Brera 29, 20121, Milano, Italy}

\begin{document}

\maketitle
\label{firstpage}

\begin{abstract} We present an analysis of star formation and feedback
recipes appropriate for galactic smoothed particle hydrodynamics
simulations. Using an isolated Milky Way-like galaxy, we constrain
these recipes based on well-established observational results.  Our
star formation recipe is based on that of Katz (1992) with the
additional inclusion of physically motivated supernova feedback
recipes.  We propose a new feedback recipe in which type II supernovae
are modelled using an analytical treatment of blastwaves.  With this
feedback mechanism and a tuning of other star formation parameters,
the star formation in our isolated Milky Way-like galaxy follows the slope and normalisation of the observed Schmidt law.
In addition, we reproduce the low density cutoff and filamentary
structure of star formation observed in disk galaxies.  Our final
recipe will enable better comparison of N-body
simulations with observations.  \end{abstract}


\section{Introduction} Meaningful comparisons between observations and
simulations of galaxies require that simulations include gas and
stars.  While dark matter controls the global dynamics of galaxies, it cannot
be directly observed.  Gas and stars provide the photons that comprise
observations.  Unfortunately, for the foreseeable future, galaxy
simulations will not be able to resolve individual stars, their
individual explosions as supernovae, or the fine structure of gas
clouds in the context of a galaxy.  It is only possible to use a
heuristic recipe to model the subresolution physics of star formation
and the feedback effect of supernova explosions.  Simulations that use
the simplest methods of injecting supernova energy into the gas
surrounding stars have proven ineffective at producing realistic star
formation feedback \citep{Katz92}.  The introduction of more clever
sub resolution schemes for the distribution of supernova energy as a
feedback on star formation (\citet{TC01} and \citet{SH03}) has resulted in more
realistic disk galaxies.

The two methods currently employed to model gas physics in simulations
are Eulerian grid codes and Lagrangian particle codes.  Eulerian codes
(e.g. \citet{CO93}, \citet{ART}, and \citet{Enzo}) track the movement of
gas around a fixed grid of cells.  The method that we describe in this
paper involves a Lagrangian treatment of the gas called smoothed
particle hydrodynamics (SPH) \citep{SPHreview}.  Resolution is
achieved naturally as particles concentrate in dense regions of
interest.  To solve the equations of hydrodynamics, physical
quantities are determined using spline kernel interpolation of
neighbouring particles.

Various methods have been employed to convert these smoothed gas
particles into stars.  Evidence of the improvement in the resolution of simulations is that early work \citep{CO93} changed dense gas
into "galaxy particles".  More recent codes \citep{Yepes97,
SwedishFeedback, SH03} choose to package stars along with hot and cold gas
into multiphase particles corresponding to the cold/warm and hot
phases of the ISM reported in \citet{MO77}.  Physical properties like
temperature and density of the multiphase particles are assumed to be
the average of the different components of each particle.  In one version of multiphase particles ('explicit'), \citet{SH03} allow gas particles to host stellar mass, spawning stars once the stellar
component of the multiphase particle exceeds a minimum star particle
mass. 

Using the SPH code GASOLINE \citep{Gasoline}, we choose to follow a
different method that is outlined in \citet{Katz92}, one that uses a
stochastic formulation for star formation (see also \citet{Kawata03}, the 'ordinary' star formation in \citet{SH03}, and \citet{Okamoto05}).  This allows stars to form
immediately and gas particles to maintain their own, unique character
and temperature, hot, cold or warm.  As supernovae are the direct
result of star formation, we use them as the feedback that limits star
formation.  The coarse mass and spatial resolution of current
simulations limits models to heuristic descriptions of this
phenomenon.  The insufficient resolution and lack of multiple gas phases also means that the $10^{51}$
ergs of energy that a supernova generates, if deposited as thermal
energy, would be dissipated through radiative cooling processes before
that thermal energy has any effect on the gas surrounding the
supernova explosion \citep[e.g.][]{Katz96, Brook04}.

Two methods have been employed to use that energy in other ways.
\citet{NW93} describe a kinematic feedback mechanism that uses the
supernova energy to provide an outwards velocity kick to all of the
gas particles surrounding a star particle in which a supernova (or
more commonly a large group of supernova) has exploded.  \citet{SH03}
use this idea to individually kick their multiphase particles in some
direction, either randomly or perpendicular to their angular momentum
vector, after a supernova has exploded. Even in such kinematic
approaches the feedback does not have a very strong effect, i.e. it is
not efficient in driving winds from small galaxies, unless the
hydrodynamic forces are temporarily turned off as in \citet{SH03}.  

An alternative to these kinematic examples is to more effectively
mimic the transfer of the kinetic energy of supernova shockwaves to
thermal energy in the interstellar medium (ISM).  \citet{Yepes97} and
\citet{SwedishFeedback} both utilise multiphase components and convert
some multiple of the mass in stars undergoing type SNII explosions
from a cold gas phase to a separate hot gas phase.  Alternatively, \citet{Pearce99} assigned gas particles to one of two phases in SPH simulations.  The phase separation can limit the loss of thermal energy deposited from supernova feedback \citep{Marri03}.  In contrast,
\citet{Gerrit97} does not explicity use multiphase particles but turns off the
radiative cooling of the gas particles immediately surrounding a star
particle in which SNII have recently exploded.  \citet{TC2000, Bottema03}
examined how the \citet{Gerrit97} scheme works in an isolated Milky
Way galaxy.  \citet{TC01, SL03, GalGov06} used this recipe in
cosmological simulations of forming galaxies.  \citet{Pelupessy04}
explored how this recipe worked in the environment of isolated disk
dwarf galaxies.  \citet{Brook04} showed that the adiabatic feedback method produces more realistic feedback in simulations of isolated collapsing halos.

We extend the exploration of the recipe started in
\citet{TC2000} for the case of an isolated Milky Way galaxy here with the
goal of further refining the star formation and feedback algorithm.
This paper specifically focuses on the star formation recipe first
described in \citet{Katz92} paired with feedback that for a time cools
only adiabatically to prevent immediate radiative losses as described
in \citet{TC2000}.  Because of this focus, we are able to optimise the
recipe, present a full parameter study and include investigations of
specific issues like the resolution dependence of star formation and
feedback.

\S\ref{sec:SF} describes the details of the recipe that we test to
convert gas into stars and \S\ref{sec:feedback} describes our feedback
schemes.  \S\ref{sec:tests} presents the tunable parameters that
affect star formation in our isolated model Milky Way (IMMW).
\S\ref{sec:results} presents the results of varying the criteria and
parameters and determines which choices work best at reproducing
observations for the IMMW.  \S\ref{sec:discussion} discusses the
relevance and plausibility of our recipe and possible future
improvements.  We present our final star formation and feedback recipe
and conclude in \S\ref{sec:conclusion}.

\section{Star Formation} 
\label{sec:SF} 
Our star formation algorithm is similar to the one proposed in \citet{Katz92} and extended in
\citet[hereafter KWH]{Katz96}. First, we apply criteria to determine
which gas particles are eligible to form stars.  We then determine
which gas particles actually form stars probabilistically such that on
average we reproduce a star formation rate formula similar to a
Schmidt law \citep{SchmidtLaw}. Those gas particles that actually form
stars spawn a new star particle of a predetermined mass, reducing
their own mass accordingly.  The new star particle is created with the
same velocity, position, and metallicity as its parent gas particle.  Star particles
can add energy, mass and metals back to gas particles through feedback
processes including type II and Ia supernova and stellar winds.

\subsection{Criteria} 
\label{sec:SFCrit} 
The recipe proposed in
\citet{Katz92} and KWH starts with an examination of every SPH gas
particle in the simulation.  The gas particle must satisfy 4 criteria
before it is eligible for star formation: 
\begin{itemize} 
\item Is the particle denser than $n_{\rm min}$ = 0.1 cm$^{-3}$?  
\item Is the particle in an overdense region?  
\item Is the particle part of a converging flow?  
\item Is the particle Jeans unstable ($\frac{h_i}{c_i} > \frac{1}{\sqrt{4\pi G\rho_i}}$)?  
\end{itemize} 
The density and overdensity criteria simply check that particles fall above the limits
set for the simulation.  We choose the overdensity limit to be 55
$\rho/\bar{\rho}$, which limits star formation to virialised regions at early
times in the Universe when the physical density everywhere is high and
plays no role in simulations of isolated galaxies. It is a simple
matter to adjust these limits to match observed galaxy properties.  In this paper we use $n$ to represent the number density (cm$^{-3}$) and $\rho$ to represent the mass density (g cm$^{-3}$).  For gas particles $n\mu m_H = \rho$, where $\mu$ is the mean molecular weight and $m_H$ is the mass of a Hydrogen atom.

\citet{Katz92} made the reasonable assumption that the gas forming a
star should be in a collapsing region and so required that the gas particles be
part of a converging flow.  In our implementation of SPH every
particle is assigned a smoothing length, $h$, such that there are a
fixed number of particles (neighbours),$N_{\rm smooth}$, within twice that
length.  We usually choose $N_{\rm smooth}$ to be 32.  Physical quantities
are estimated using spline kernel interpolation. For example, the
mass density, $\rho$, for particle $i$ is given by \begin{equation}
\rho_{\rm i} = \sum_{\rm j=1}^{N} m_j W(|\textbf{r}_i-\textbf{r}_j|,h_i,h_j)
\end{equation} where $m$ is the particle mass, N is the number of gas
particles, and $W$ is the smoothing kernel, which we choose to have
compact support, i.e. it goes to zero beyond $2h$ so the sum is really
only over $N_{\rm smooth}$ \citep{SPHreview,HK89}.  The divergence of the
velocity field, $\nabla\cdot\textbf{v}$ at the position of gas
particle $i$ is given by \begin{equation} \nabla\cdot\textbf{v} =
{1\over \rho_i}\sum_{\rm j=1}^{N} m_j(\textbf{v}_j-
\textbf{v}_i)\cdot\nabla_i W(|\textbf{r}_i-\textbf{r}_j|,h_i,h_j)
\end{equation} where {\textbf v} is the velocity.  When
$\nabla\cdot\textbf{v}$ is negative, the criterion is satisfied, it is
assumed that the gas particle is part of a collapsing flow, and the
gas particle can form stars.

The Jeans Criterion is a test of whether or not a gas cloud can
provide pressure support against gravitational collapse.  If a sound
wave cannot travel across the cloud in the time it would take the
cloud to gravitationally free fall to the centre, then the cloud will
collapse.  \citet{Katz92} proposed that such a criterion should take
the form: \begin{equation} \frac{h_i}{c_i} > \frac{1}{\sqrt{4\pi
G\rho_i}} \end{equation} where $c_i$ is the sound speed of the gas
particle in question and $G$ is the gravitational constant. This tests
if the temperature and pressure of one particle inside its smoothing
sphere would be able to support the whole sphere against gravitational
collapse.  \citet{Katz92} only applies this criterion when the region
is not affected by gravitational softening, but \citet{Okamoto03} notes that the Jeans Criterion formulated in this way is similar to the \citet{Bate+Burkert97} resolution limit for artificial fragmentation.  In \S \ref{sec:SFCrit}, it is shown that this formulation of the Jeans criterion introduces a resolution
dependence and hence we will eliminate it from our final formulation.

\subsection{Stochastic Star Formation} \label{subsec: stoch} Ideally,
stars should form whenever specific criteria like those described
above are met (see also \citet{Li04} or \citet{M-D04}).  However, since our simulations
of galaxies have limited resolution we can only capture the global
behaviour of star formation, and we use a probabilistic approach.

\citet{Katz92} bases the number of stars that form on the theoretical
star formation rates of \citet{Larson69} and \citet{Silk87}, who both
proposed that $\rho_{\rm SFR} \sim \rho_{\rm gas}^{3/2}$, where $\rho$ is the
volume density.  These formulations make use of the fact that the
dynamical time, $t_{\rm dyn} \sim \rho^{-1/2}$.  However, to form stars
gas must also cool so \citet{Katz92} chose the star formation
timescale, $t_{\rm form}$, to be the maximum of the dynamical time and the
cooling time. If the gas was already cool enough to form stars,
i.e. $T<T_{\rm max}$, then $t_{\rm dyn}$ was used. \citet{Katz92} took
$T_{\rm max}=30,000$K, which was appropriate given his cooling
function. So one can write the star formation rate as \begin{equation}
\frac{d\rho_{\rm \star}}{dt} = c^{\star} \frac{\rho_{\rm gas}}{t_{\rm form}}
\label{eq:starform} \end{equation} where we have introduced a constant
efficiency factor $c^{\star}$ that will enable us to adjust the star
formation rate to match observations.  We will choose to set $t_{\rm form}
= t_{\rm dyn}$ at all times and instead will introduce another star
formation criterion that $T < T_{\rm max}$.

Because of the dependence on density, it is possible to create a
stochastic recipe for when and where stars should form.  If one takes
the probability that a star will form as \begin{equation} p =
{m_{\rm gas}\over m_{\rm star}}\left(1 - e^{-c^{\star}\Delta
t/t_{\rm form}}\right) \label{eq:prob} \end{equation} where $m_{\rm 
gas}$ is
the mass of the gas particle and $m_{\rm star}$ is the mass of the
potentially spawned star particle then on average one recovers
Equation (\ref{eq:starform}).

\begin{figure} \begin{center} \resizebox{9cm}{!}{\resizebox{9cm}{!}{\includegraphics{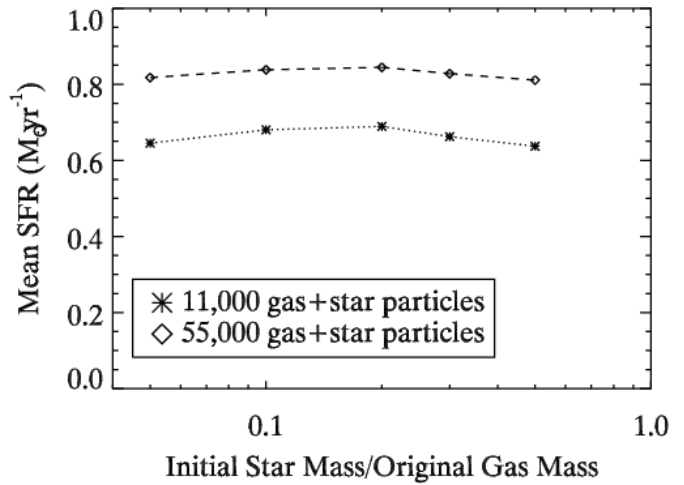}}}
\caption{ The mean star formation rate (SFR) as a function of the star
particle mass in units of the original gas particle mass.  The SFR
shows little dependence on the initial star particle mass.  These
results are for our fiducial values of parameters and criteria on the
isolated model Milky Way (IMMW) described in \S\ref{sec:tests}.}
\label{fig:badismass} \end{center} \end{figure}

In this formulation there is a greater probability of a star forming
in denser areas.  For each star formation eligible gas particle a
random number, $r$, is drawn between zero and one and if $r<p$ a new
star particle is created.  \citet{Katz92} assumed that the star
particle mass was always a constant fraction of the parent gas
particle mass, i.e.  $m_{\rm star}/m_{\rm gas} = \epsilon^\star$.  In 
that case, $\epsilon^{\star -1}$ replaces $m_{\rm gas}/m_{\rm star}$ in Equation
(\ref{eq:prob}). $\epsilon^\star$ can be absorbed into the definition of
$c^\star$, for the relevant case where $\Delta t/t_{form}$ is small, and hence
does not appear in Equation (5) of \citet{Katz92}. Therefore, to compare with
\citet{Katz92}, $c^\star$'s here should be multiplied by $\epsilon^\star$.
Here we are interested in
making the new star particle have a constant mass to remove the large
variation in star particle mass that can occur as stars form at
different times during the simulation.  Using Equation (\ref{eq:prob})
and assuming a constant stellar particle mass we can recover the same
global star formation rate, almost independent of that mass, as shown
in Figure \ref{fig:badismass}. A smaller star particle mass gives us
better resolution for the stellar component at the cost of greater
computational expense and we choose a value of 0.2$\times$ the initial
gas particle mass as a compromise.

Codes that use multiphase gas particles for star formation, e.g.
\citet{SH03} 'explicit' star formation, more finely time sample star formation since a specific
amount of gas is converted into stars during every gas particle
timestep and new star particles are only spawned periodically.
However, this approach has an inherent problem. When feedback occurs,
stars that are still tied to their parent gas particles can get
hydrodynamically accelerated and possibly even be ejected from the
galaxy, which is not physical.

One problem that could arise when using a stochastic approach for star
formation is that particularly dense regions might not experience the
vigorous star formation that is observed in starburst galaxies
\citep{Telesco88, Chapman03} if the mass of all star particles is
fixed.  A possible solution to this problem is to make the mass of the
star particle that forms dependent on properties of the gas particle.
For instance, \citet{E+E97} provide a formula using the density and
the pressure of gas to determine what fraction of the gas will form
stars.  Initial tests using this formula suggest that our resolution
is too coarse for this recipe to yield useful results.  However, for
our fiducial parameters, a star particle mass 0.2 times the
initial gas particle mass and a star formation interval, $\Delta t$,
of 1 Myr, this problem would only arise if gas consumption
times become less than 5 Myr, a situation that is rarely, if
ever, observed.

\section{Feedback} 
\label{sec:feedback} 
\citet{Katz92} and KWH add
mass and energy feedback from type II supernova.  The energy is added
gradually with an exponential decay rate of 20 Myr. It is
added at the location of the parent gas particle and is smoothed using
the SPH smoothing kernel.  Since this thermal energy is typically
added to very dense gas, it is quickly radiated away and has little
effect on the evolution of the galaxy.  We describe a stronger and
perhaps more realistic method for including type II supernovae (SNII)
feedback \citep{Gerrit97}, which we implement here.  In addition we
now also include feedback from type Ia supernovae and stellar winds
from planetary nebulae and allow the metals produced in our stars to
be distributed.  Their implementation is described towards the
end of this section.

\subsection{Type II Supernovae (SNII)} SNII play a dominant role in
regulating star formation in our simulations because of their ability
to heat large volumes of the interstellar medium near the site of star
formation and consequently prevent more gas from collapsing
\citep{Silk03}.  Because the blastwave shocks of SNII convert the
kinetic energy of ejecta into thermal energy on scales smaller than
our simulations resolve, feedback in our simulations is purely
thermal.

The number of supernovae produced by a star particle depends on the
initial mass function of the stars that form.  We use the three piece
power law fit of the IMF defined in \citet{MS79}, where $\alpha$ =
-0.4 for stars with masses between 0.1 and 1 $M_{\rm \odot}$; $\alpha$ =
-1.5 for stars with masses between 1 and 10 $M_{\rm \odot}$; $\alpha$ =
-2.3 for stars with masses greater than 10 $M_{\rm \odot}$.  Our IMF
starts at 0.1 $M_{\rm \odot}$ because \citet{RHG} report that the stellar
luminosity function appears to turn over at a luminosity corresponding
to that mass, so that stars with a mass $<$ 0.1 $M_\odot$ do not make
a large contribution to the total stellar mass.

The time when a star explodes as a supernova depends on its
lifetime. We use the \citet{Raiteri96} parameterisation of the Padova
group's \citep{Alongi93,Bressan93,Bertelli94} stellar lifetime
calculations for stars of varying metallicities.  In this
parameterisation, since more massive stars have shorter lifetimes than
small mass stars, it is possible to determine the maximum and minimum
stellar mass that will explode during a given timestep and, therefore,
integrating over the initial mass function provides the total mass and
number of stars that will explode.  Like the \citet{Raiteri96} recipe,
we only allow stars between 8 and 40 $M_{\rm \odot}$ to explode as SNII;
stars more massive than this are assumed to either collapse into black
holes or explode as type Ib supernovae.  Regardless, few stars form
with masses greater than 40 $M_{\rm \odot}$, so the impact on the feedback
is minimal.  The use of stellar lifetimes is an improvement on
previous SN feedback recipes that bled the supernova energy out
gradually as some type of exponential function after a star particle
formed \citep[KWH]{CO93}.  Thus, many implementations of feedback choose to use them now \citep{Lia02,Kawata03, Okamoto03, Scan05}.

We multiply the number of SNII that explode by the energy ejected into
the ISM, $E_{\rm SN}$, a fixed fraction of the canonical $10^{51}$ ergs
produced by a supernovae, and distribute that energy to the
surrounding gas particles.  The energy is spread out, weighted by the
mass of the gas particle that receives it, using the SPH smoothing
kernel. Unlike in \citet{Katz92} and KWH, however, the feedback is
centred on the current position of the star particle and not on the
parent gas particle. Therefore, the feedback energy received by a
neighbouring gas particle $i$ for a total feedback energy of $\Delta
E_{\rm SN}$ is just 
\begin{equation} \Delta E_{\rm SN,i} = {m_i
W(|\textbf{r}_i-\textbf{r}_s|,h_s) \Delta E_{\rm SN}\over \sum_{\rm j=1}^{N}
m_j W(|\textbf{r}_j-\textbf{r}_s|,h_s)} 
\label{eq:smooth}
\end{equation} 
where $h_s$ is the distance from the star particle to
its 32nd closest neighbouring gas particle.  We explore how star
formation depends on $E_{\rm SN}$ in \S\ref{sec:results}.  When $E_{\rm SN}$ =
$10^{50}$ ergs, $7.65\times10^{47}$ ergs of energy are
deposited into the surrounding gas for every one $M_{\rm \odot}$ of stars
formed.  Metal production follows \citet{Raiteri96}, where the ejected
oxygen and iron mass are estimated using a power law based on the mass
that \citet{Raiteri96} estimated from \citet{Woosley95}.  We integrate
this power law between the minimum and maximum mass stars that explode
during a given timestep to determine the mass of oxygen and iron
ejected into the ISM.  The metals and the mass returned to the gas
particles by the type II supernova are distributed in a way similar to
Equation (\ref{eq:smooth}).

Unfortunately, the distributed feedback energy will have little impact
on our simulated galaxies owing to our finite resolution and inability
to resolve the complex multiphase interstellar medium properly.  Since
the supernovae explode in regions of high average density, the gas can
radiate away the energy in much less time than a typical timestep.  To
make the feedback more realistic we tried two different approaches
(see \S \ref{sec:betamodel} and \S \ref{sec:bwrecipe}) to mimic
blastwaves, which should take hundreds of thousands of years to cool.
In both approaches we disable the radiative cooling in a number of the
nearest gas particles.

Disabling the radiative cooling in the surrounding gas particles
allows us to model two different aspects of the feedback phenomenon.
First, the gas particles immediately surrounding new stars have their cooling disabled, and with the added energy from the supernovae, will likely become hotter
than the maximum temperature, $T_{\rm max}$, allowed for forming stars.
In this way, feedback inhibits further star formation in dense regions much like
supernovae generate turbulence in molecular clouds, which provides
global stability against further collapse of the entire molecular
cloud.  Secondly, the increased temperature of the gas particle models
the high pressure of a blastwave, which plays a key role in shaping
the interstellar medium, allowing the surrounding gas to naturally flow
outwards.

In our initial scheme, we determine the number of gas particles that
have their cooling disabled by multiplying the SNII mass by a mass
loading factor, $\beta$, and shutoff the cooling for a constant time,
$\tau_{\rm cso}$.  Our other scheme requires fewer parameters because it
depends on the analytic treatment of blastwaves described in
\citet{MO77}.

Both schemes allow gas particles to receive energy from multiple supernovae explosions in a similar fashion.  Every timestep, each gas particle has its cooling shutoff time, $\tau_{\rm CSO}$ calculated from the total supernova energy received.  For subsequent supernova, $\tau_{\rm CSO}$ is recalculated and extended when necessary.

\subsubsection{Supernova Mass Factor ($\beta$) Recipe}
\label{sec:betamodel} We derive the mass loading factor concept from
multi-phase recipes like \citet{Yepes97} and \citet{SwedishFeedback}.
We calculate the exact number of gas particles where we turn off the
radiative cooling in the following manner.  For gas particles
surrounding a star particle, radiative cooling is disabled for gas
particles within a sphere of mass $\beta M_{\rm SNII}$.  For a gas particle $i$, if

\begin{equation} \beta M_{\rm SNII} > \frac{4 \pi
(|\textbf{r}_i-\textbf{r}_s|)^3}{3} \rho_{\rm ave} \end{equation} 
where $M_{\rm SNII}$ is the mass of stars that go supernovae during
a given timestep, $(|\textbf{r}_i-\textbf{r}_s|)$ is the distance from
the star to the gas particle in question, and $\rho_{\rm ave}$ is given by
\begin{equation} \rho_{\rm ave} = \sum_{\rm j=1}^{N} m_j
W(|\textbf{r}_s-\textbf{r}_i|,h_s).  \end{equation} 
The maximum number of particles for which we can disable the cooling is $N_{\rm smooth}$,
which is 32 in our simulations.

In our initial recipe, cooling is disabled for a fixed amount of time,
$\tau_{\rm CSO}$.  We started with $\tau_{\rm CSO}$= 30 Myr based on the work
of \citet{Gerrit97} and \citet{TC2000} who suggest that 8 $M_\odot$
stars have a lifetime of 30 Myr (although using the Padova tracks it
would be closer to 38 Myr). Thus, after 30 Myr, feedback produced in a
star forming region should be finished.  We explore the effects of
varying $\tau_{\rm CSO}$ in \S\ref{sec:taucso}.

\subsubsection{Blastwave Recipe} \label{sec:bwrecipe} Exploration of
the $\beta$ parameter motivated us to introduce an explicit blastwave
solution based on \citet{Chev74} and \citet{MO77}.  This solution
reduces the number of tunable parameters by providing both the maximum
radius to which the blastwave explosion will reach and the time that
the blastwave will keep the gas hot.  The maximum radius of a
supernova blastwave in the \citet{Chev74} simulations was
\begin{equation} R_E =
10^{1.74}E_{\rm 51}^{0.32}n_0^{-0.16}\tilde{P}_{\rm 04}^{-0.20} {\rm pc}
\end{equation} where $E_{\rm SN}=E_{\rm 51} 10^{51}$ ergs, $n_0$ is the
ambient Hydrogen density, $\tilde{P}_{\rm 04} = 10^{-4}P_0 k^{-1}$ where $P_0$ is
the ambient pressure and k is the Boltzmann constant.  Both $n_0$ and $P_0$ are calculated using the SPH kernel for the gas particles surrounding the star.  We temporarily turn off the radiative cooling
for all gas particles within $R_E$.  However, there is an artificial maximum of 32 particles that can have their cooling disabled.  Figure \ref{fig:bwstats} shows how many of these particles there are typically.

The simulations also provide a timescale for the time that a gas
particle does not radiatively cool.  Naively, one might think that
this timescale should be the length of the Sedov phase of a supernova
explosion.  During this phase, energy is conserved in the supernova
remnant because it is not able to effectively radiate.  However, the
Sedov phase only lasts for tens of thousands of years \citep{PadBook}.
Our simulations cannot resolve this timescale and the feedback would
be ineffective even if our simulations produce clusters of supernova.
Also, \citet{MO77} suggest that a hot, low density shell survives well
after the Sedov phase.

Following the Sedov phase, blastwaves enter the snowplow phase.
During the snowplow phase, momentum is conserved as the blastwave
expands because the gas has cooled enough to radiate more efficiently.
\citet{MO77} present the end of the snowplow phase when a supernova
remnant first reaches its maximum extent: 
\begin{equation} t_{\rm E} =
10^{5.92}E_{\rm 51}^{0.31}n_0^{0.27}\tilde{P}_{\rm 04}^{-0.64} {\rm yr}
\end{equation} The supernova remnant continues to cool radiatively
even after it stops expanding.  \citet{MO77} report that the time that
the hot, low density shell will survive is 
\begin{equation} t_{\rm max} =
10^{6.85}E_{\rm 51}^{0.32}n_0^{0.34}\tilde{P}_{\rm 04}^{-0.70} {\rm yr}
\end{equation} 
Either of these timescales may be appropriate for the
length of time to disable cooling and we report on how using either
$t_E$ or $t_{\rm max}$ affects star formation in \S\ref{sec:largeSN}.

Even in the case where a supernova has exploded in a previous timestep, we currently use the previous equations as described.  We are investigating the possible interactions of multiple supernovae remnants.  As an initial assumption, we suppose that all of the supernovae exploding during a given timestep combine their energy to generate the blastwave.

\subsubsection{Small SN Smoothing} 
Since we only disable cooling for a fraction of the particles within the smoothing radius, it is only
those particles that maintain the high temperature generated from the
supernova.  Thus, all the energy that gets distributed beyond the
blast radius is quickly radiated away, which is still unphysical.  To
address this problem, we introduce another variant of the blastwave
approach where we restrict the distribution of energy from the supernova only to those particles within the blast radius using a kernel function limited to just the particles within the blast radius.  Initial trials only distributed metals and mass inside the blast radius like the energy.  However, as we have yet to implement diffusion of metals between gas particles, the supernova explosion represents the only time when metals can be widely distributed.  Distributing metals only inside the blast radius lead to spurious metal distributions.  Thus, we reverted to distributing the metals and mass across the entire smoothing sphere.

It sometimes occurs that no particle is within the blast
radius.  In this case, we deposit the energy, metals, and mass to the
nearest gas particle. Ejecta are distributed in this manner for both
SNII and SNIa, but for SNIa, the cooling is not disabled.  If there
are no supernova ejecta, the wind feedback is distributed across all
32 nearest neighbours with the standard smoothing radius.  

Such an approach might be a cause for concern
since \citet{Benz90} have found that depositing energy into a single
gas particle in a SPH simulation can lead to overcooling and a
violation of energy conservation owing to the large energy gradient
introduced. However, \citet{SpringelEntropy} show that this is not a
problem if one uses the asymmetric form of the thermal energy
equation, as we do in GASOLINE.

\subsection{Type Ia Supernovae (SN Ia)} SNIa are also significant
sources of metals and are thought to occur in binary systems.  The
method we use to determine how many SNIa explode is again described in
detail in \citet{Raiteri96}.  The minimum mass of a binary system is 3
$M_\odot$ and the maximum mass is 16 $M_\odot$ (two 8 $M_\odot$
stars).  Using the binary fractions from \citet{Raiteri96} makes the
number of SNIa 10-20 $\%$ of the total supernovae in our Isolated Model Milky Way simulations,
as observed by \citet{vdBMcC94} in spiral galaxies.

Usually we distribute the SNIa energy using the smoothing kernel
amongst the $N_{\rm smooth}$ nearest neighbouring gas particles.  However,
for those star particles that also distribute SNII energy within the
blast radius, the SNIa energy is also only distributed within the
blast radius.

Radiative cooling is not disabled as a result of SN Ia because SN Ia
occur much after the stars initially form.  During this time, stars
would dynamically spread out of their initial associations and
subsequent SNIa would not be a collective phenomenon, like SNII, and
hence would not lead to large blastwaves.  Since stars in our
simulations consist of indivisible particles, the SNIa stars cannot
drift apart as they should, would act collectively and produce too
large an effect if we turned off the cooling.  Our early simulations indicated the inclusion of SNIa produced too large of a feedback effect.

Like energy, mass and metals are smoothed across all $N_{\rm smooth}$ nearest neighbour
particles.  All SNIa are assumed to eject the same mass (1.4
$M_\odot$) and the same amount of iron (0.63 $M_\odot$) and oxygen
(0.13 $M_\odot$) based on the \citet{SNIaYields} SNIa yield models.
These quantities are added to the existing oxygen and iron in the gas
particles by mass and then converted to a fractional mass of the gas
particle so that when a new star forms, it will form with the same
fractional metal content as its parent gas particle.

\subsection{Stellar wind feedback}

The feedback contribution of stellar winds is also significant.  Stars
with masses below $\sim$8 M$_\odot$ return substantial fractions of
their mass to the ISM as they evolve and leave behind white dwarf
remnants.  We base our wind feedback on the work of
\citet{kennicutt94} who find that the total stellar return fraction is
0.25 to 0.50 of the initial mass depending on the IMF.  Because the
return rate is so high, this form of feedback can greatly prolong star
formation in galaxies without gas inflow.

For simplicity, we consider only stars between 1 and 8 M$_\odot$ and
assume that lower mass stars remain unevolved.  To determine the
fraction of mass returned for a given stellar mass we use the
initial-final mass relation of \citet{weidemann87} and then fit his
results to a continuous function.

In practical terms, we implement this feedback mechanism by first
taking each star particle and determining the range of stellar masses
that die during the current timestep using the lifetimes from
\citet{Raiteri96}.  Then we calculate a returned mass fraction for
this mass range using the function derived from \citet{weidemann87}.
We add the feedback to the gas particles in the same manner as the SNe
feedbacks, except without injecting any energy.  The metallicity of
the returned gas is simply the metallicity of the star particle. In
the future, we plan to include metal production by intermediate
mass stars.  The total fraction of mass lost from a star particle over
$>$10 Gyr is 40\% and of this $\sim$99\% of the mass loss results from
stellar winds.

\section{Tests of the Recipe} \label{sec:tests} To closely examine our
star formation formalism, we created an isolated model Milky Way
(hereafter IMMW) based on the dynamical model presented in
\citet{KZS02} at several different resolutions.  We use these models
to tune our star formation recipe to produce results consistent with
observations \citep[e.g.][]{LocalSFR, Kenn98, Wong+Blitz02}.

\subsection{Isolated Galaxy}

The IMMW was created using the specifications of \cite{Springel00} in
that it resides in a slightly modified \cite{NFW} (hereafter NFW) dark
matter halo where the central dark matter has been concentrated by
infalling baryons.  Thus, the density distribution and potential is
slightly different from a pure NFW halo.  We initially modelled the
dark matter using a velocity $v_{\rm 200}$=150 km s$^{-1}$, a concentration, $c =
r_{\rm 200}/r_s=12$, resulting in a mass $M_{\rm 200} = 1.12 \times 10^{12}
M_{\rm \odot}$ and radius $r_{\rm 200}$=214 kpc at an average overdensity
$\frac{\delta \rho}{\rho_{\rm c}}$ = 200.  However, to replicate as
closely as possible the potential created in the \citet{Springel00}
model, we fit a slightly different NFW profile to the particle
potential.  This fixed potential has $M_{\rm 200} = 7.314 \times 10^{11}
M_{\rm \odot}$, $r_{\rm 200}$=153 kpc, and concentration c=20.648.

The baryons are distributed in a stellar and gaseous disk along with a
spherical bulge that contains only star particles with a total
baryonic mass of 4.55$\times10^{10} M_{\rm \odot}$.  The bulge contains
4.93$\times10^{9} M_{\rm \odot}$, $\sim10$\% the baryonic mass.  The stellar
disk follows an exponential profile with a scale length of 3.5 kpc and
constitutes 90$\%$ of the total mass of the disk.  We distributed the
remaining 10$\%$ of the baryonic disk mass as collisional SPH gas
particles in an exponential profile with a scale length of 7 kpc,
twice the scale length of the stellar disk \citep{Broeils+Rhee97}.  We
introduce no holes or gaps in the gas because molecular gas studies
show that where neutral atomic hydrogen column densities decrease in
the inner parts of disks, molecular gas densities increase to fill the
void \citep{Wong+Blitz02}.

We evolve the disk until the initial instabilities (caused by the
initially smooth particle distribution developing a spiral pattern)
die out before turning on star formation.  The minimum Toomre Q value
is 2, so this disk is a very stable disk and grows no bar.

To check whether our star formation rate converges to a single value
at various resolutions, we replicate our model three times.  The
lowest resolution galaxy starts with only 9,000 star particles of
5$\times10^{6} M_{\rm \odot}$ and 2,000 gas particles of 2.5$\times10^{6}
M_{\rm \odot}$, the medium resolution galaxy has 45,000 star particles of
1$\times10^{6} M_{\rm \odot}$ and 10,000 gas particles of 5$\times10^{5}
M_{\rm \odot}$, and the highest resolution galaxy has 225,000 star
particles of 2$\times10^{5} M_{\rm \odot}$ and 50,000 gas particles of
1$\times10^{5} M_{\rm \odot}$.  We soften the gravity using spline
softening and our gravitational softening length, 650 pc for the gas
particles and 325 pc for star particles, remains fixed for all three
resolutions. The equivalent Plummer softening is about 0.7 times
smaller.

\subsection{Goals of IMMW} 
Many observations have been dedicated to
studying star formation in the Milky Way and spiral galaxies similar
to our IMMW.  The observations allow us to constrain the values of our
star formation recipe parameters.  \citet{SchmidtLaw} provides the
foremost constraint for how many stars should form in a dense gas
environment.  Schmidt showed that the surface density of star
formation, $\Sigma_{\rm SFR}$, follows a power law of the gas surface
density, $\Sigma_{\rm gas}$, called the Schmidt law.  The more recent work
of \citet{Kenn98} specifies the exact slope and normalisation of this
relationship.  Equation 4 of \citet{Kenn98} states that:
\begin{equation} \label{eq:Kenn} \Sigma_{\rm SFR} = (2.5 \pm 0.7) \times
10^{-4} (\frac{\Sigma_{\rm gas}}{1 M_{\rm \odot} {\rm pc^{-2}}})^{1.4\pm0.15}
M_{\rm \odot} {\rm yr^{-1} kpc^{-2}} \end{equation} 
The formulation of our star
formation recipe should ensure that our star formation approximately
follows the slope of this relationship, while the star formation
efficiency, $c^\star$, should adjust the normalisation.

Another constraint is the observed nearly steady star formation rate
in the Milky Way.  The local stellar neighbourhood shows evidence that
the star formation rate has been constant for Gyr \citep{LocalSFR} when averaged over long timescales.  In our simulations, a steady star formation rate results from the
maintenance of a constant exponential gas surface density profile.
The \citet{Wong+Blitz02} observations of exponential gas surface
density profiles, therefore, provide another constraint related to
the steady star formation rate.

Our experiments consist of varying the four star formation criteria
(temperature, density, converging flow, and Jeans) and the four
parameters ($c^\star, E_{\rm SN}, \beta$, and $\tau_{\rm CSO}$) to determine
how each criterion and each parameter affects star formation.
Criteria are solid cutoffs that eliminate gas particles from forming
stars whereas parameters are proportional constants that affect the
rate of star formation and feedback.  In this spirit, we choose a
fiducial set of criteria and parameters, and then proceed to vary each
parameter or criterion individually.  Our fiducial criteria are
$T_{\rm max}$ = 30,000 K, $n_{\rm min}$ = 0.1 cm$^{-3}$, flows must be
converging, and {\emph no} Jeans-like criterion.  The fiducial
parameters are $\beta$=10,000, $c^\star$ = 0.1, $\tau_{\rm CSO}$ = 3
$\times 10^7$ yr, and $E_{\rm SN}$ = $10^{50}$ ergs.  These
parameters do not necessarily represent a best fit, which is presented
in the conclusions, but are simply a starting point that produce
relatively normal results.

\subsection{Numerical Precision} 
Our star formation recipe is
implemented in the parallel tree SPH code GASOLINE \citep{Gasoline}.  GASOLINE implements cooling similar to what is described in KWH.  It assumes ionisation equilibrium, an ideal gas with primordial composition, and solves for the abundances of each ion species.  The scheme uses the collisional ionisation rates reported in \citet{Abel97}, the radiative recombination rates from \citet{Black81} and \citet{Verner96}, bremsstrahlung, and line cooling from \citet{Cen92}. The energy integration uses a semi-implicit stiff integrator
independently for each particle with the compressive heating and density
(i.e. terms dependent on other particles) assumed to be constant over the
timestep.  

Gravity is calculated for each particle using tree elements that span
at most $\theta$ = 0.7 of the size of the tree element's distance from
the particle.  Every particle has its forces calculated on each large
time-step, 1.53 $\times 10^7$ yr.  GASOLINE is multistepping so
that every particle's time-step
$\Delta{t_{\rm grav}}=\eta\sqrt{\frac{\epsilon_i}{a_i}}$, where $\eta$ =
0.175, $\epsilon_i$ is the gravitational softening length, and $a_i$
is the acceleration.  For gas particles, the time-step must also be
less than $\Delta{t_{\rm gas}}=\eta_{\rm Courant}\frac{h_i}{c_i}$, where
$\eta_{\rm Courant}$ = 0.4 and $c_i$ is the sound speed. We restricted the
smallest SPH smoothing length to be 0.01 of the gravitational
softening length.

Stars are formed and feedback is calculated every 1 Myr in
the simulations.  The time between star formation events has no
relationship to the major timesteps of the simulation when every
particle has its forces calculated.  However, star formation is tied
to the minor timesteps when some subset of the particles have their
forces calculated.  As these timesteps may not be exactly 1 Myr, we choose the minor timestep that is closest to the time that
we want to form stars and add feedback at that time.

\section{Results} 
\label{sec:results}
\begin{figure} 
 \begin{center}
\resizebox{9cm}{!}{\resizebox{9cm}{!}{\includegraphics{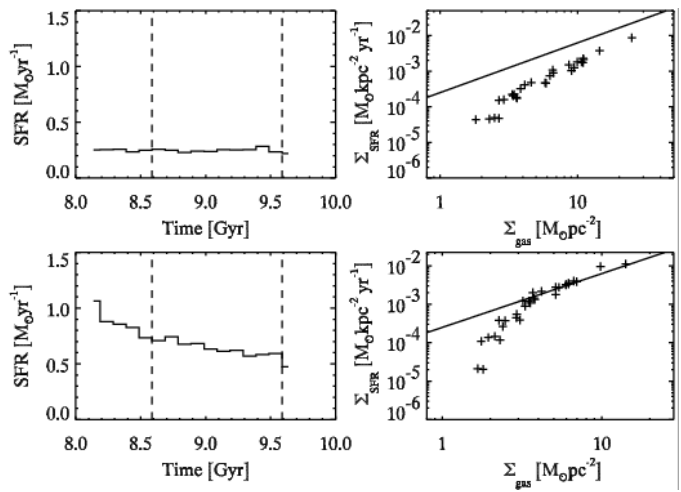}}} 
\caption{A demonstration of unacceptable
(top) and acceptable (bottom) star formation conditions.  The top panels are the results produced when using a small star formation efficiency, $c^\star$ of 0.01 in contrast to
the bottom panels that use the best fit value of $c^\star$=0.05 in the blastwave model.
The left panels plot the SFR vs. time and the right panels the SFR
surface density vs. gas surface density relation. The two dashed lines in the
left panel show the region that is used to calculate the average
SFRs. The solid line in the right panel is the observed Schmidt Law
\citep{Kenn98}.  } \label{fig:goodbad} \end{center} \end{figure}

We have conducted a variety of experiments with our
IMMW, adjusting the parameters discussed above.  For each of these
experiments, we have data on the star formation rate (SFR) as a
function of time and gas surface
density\footnote{http://hpcc.astro.washington.edu/feedback}.
In Figure \ref{fig:goodbad}, we present two parameter choices and show
star formation histories and SFR surface density versus gas surface density
relations as an example of star formation variation within our
experiments.  In this case, the blastwave feedback model is used in medium resolution (45,000 stars, 10,000 gas) IMMWs.  The top panels are the results produced when using a
small star formation efficiency, $c^\star$ of 0.01 in contrast to
 the best fit for the blastwave model $c^\star=$0.05 displayed in bottom plots.

The star formation history, plotted in the left panels, is a histogram
of when stars formed. The vertical dashed lines indicate the period
between 6$\times10^8$ and 1.6$\times10^9$ yr after the beginning of
the simulation, the time over which we calculate the average SFR,
which we use in subsequent plots.  The simulations start 8 Gyr after the stars in the initial conditions formed so that there are no feedback effects from those stars.  We choose a time period past the
beginning of the simulation because it falls well after any initial
transient starburst that may result from non-equilibrium initial
conditions.  Such a starburst can happen because all the gas particles
at the start of the simulation are unaffected by feedback and hence
all of them may be eligible to form stars.  As shown in the panels,
both choices result in an approximately constant star formation
history but the amount of present day star formation in the upper
panel is lower than the observed value.

The right panels of Figure \ref{fig:goodbad} show SFR surface density
versus gas surface density and indicates how well our star formation
follows a Schmidt law \citep{Kenn98}.  To create these plots, we
azimuthally sum the mass of stars that formed over the last 100
Myr of the simulation in 500 pc radial bins and plot them
against the final surface density of the gas in each bin.  The panels
show that our star formation formulation creates a Schmidt Law with
the right slope. However, the upper panel does not have the correct
amplitude suggesting that the choice of parameters was incorrect.  It
is generally the case that an average star formation rate of around
0.8 $M_\odot$ yr$^{-1}$ in the high resolution case reproduces the Schmidt
Law in the IMMW experiments.  The turn down in SFR density at
surface densities less than 2 $M_\odot$ pc$^{-2}$ matches the observations of \citet{MK01} as we discuss in Section \ref{sec:sum}.

\begin{figure} 
 \begin{center}
\resizebox{9cm}{!}{\resizebox{9cm}{!}{\includegraphics{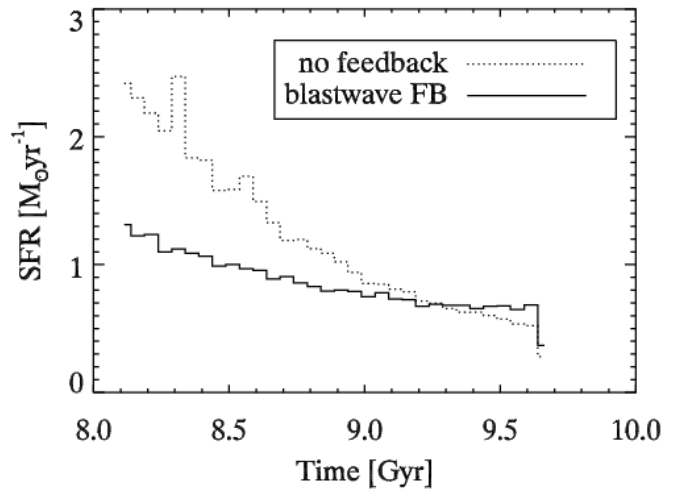}}} 
\caption{ A comparison between star formation with ({\emph solid}) and without ({\emph dotted}) feedback.  The run with feedback uses $E_{\rm SN} = 3\times 10^{50}$ ergs with the blastwave model at high resolution (275,000 particles).   } \label{fig:nofbsfr} \end{center} \end{figure}

\begin{figure} 
 \begin{center}
\resizebox{9cm}{!}{\resizebox{9cm}{!}{\includegraphics{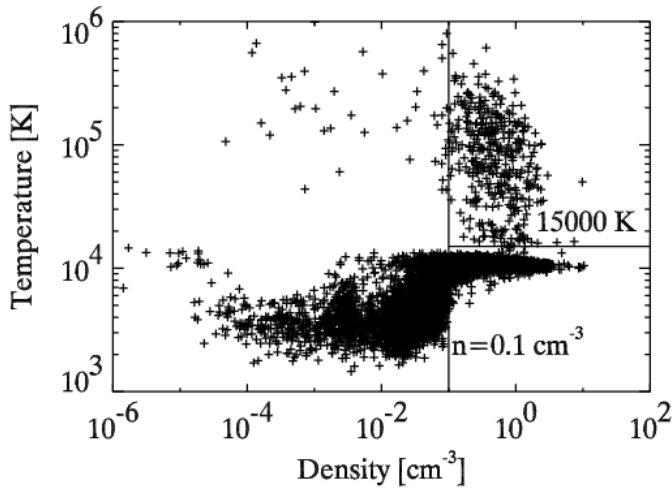}}} 
\caption{ The phase diagram for the 50,000 gas particles in the high resolution (275,000 paticles) IMMW run with $E_{\rm SN} = 3\times 10^{50}$ ergs using the blastwave model.  The fiducial values for $T_{\rm max}$ and $n_{\rm min}$ are drawn to indicate the gas particles that pass the star formation criteria.  All of the particles with temperatures above 15,000 K are there as a result of feedback because all gas particles in the disk start with T=10,000 K.  A couple of gas particles show a modest decrease in density below the density threshold indicating that the gas does expand slightly owing to its high temperature and pressure.  } \label{fig: phasediag} \end{center} \end{figure}

The effect of feedback is previewed in Figure \ref{fig:nofbsfr} from a high resolution (225,000 stars, 50,000 gas) IMMW.  The shallower drop-off in star formation rate with feedback owes in part to the fact that the supernova energy is being effectively used to prevent star formation and in part due to the stellar wind feedback that returns gas to the ISM.  Both star formation rates decline exponentially because of the stochastic recipe (\S \ref{subsec: stoch}).  Figure \ref{fig: phasediag} shows how the feedback immediately increases the temperature of the gas surrounding stars that have recently formed.  The higher temperature results in higher pressure that allows the hot particle to push around the cooler gas surrounding it, as shown in Figure \ref{fig:gaspic}.  The expansion leads to only a modest reduction in density because density is a smoothed property that includes the surrounding high density particles.  Thus, it is not low gas density that suppresses star formation, but high gas temperature.

\subsection{Effects of Criteria} \label{sec:crit}

\subsubsection{$T_{\rm max}$: Maximum Temperature} \label{sec:Tmax} As we
already stated, we added an additional criterion to those in
\citet{Katz92} and KWH: gas particles may not form stars unless their
temperature is below $T_{\rm max}$, typically 10,000's K. This may seem
like a high temperature threshold for star formation given that star
forming molecular clouds are observed to cool down to $\sim 100$ K.
However, our cooling is limited to H and He atomic cooling, which can
only cool gas down to $\sim 10,000$ K, and we average over scales much larger
than star forming clouds.  A future improvement to the code will be to
include molecular hydrogen cooling (e.g. \citet{Abel97} and
\citet{Kravtsov03}), which will allow the gas to cool below 10,000 K,
but even then, unless the resolution were greatly improved, $T_{\rm max}$
should remain above 10,000 K.

\begin{figure} \begin{center} \resizebox{9cm}{!}{\resizebox{9cm}{!}{\includegraphics{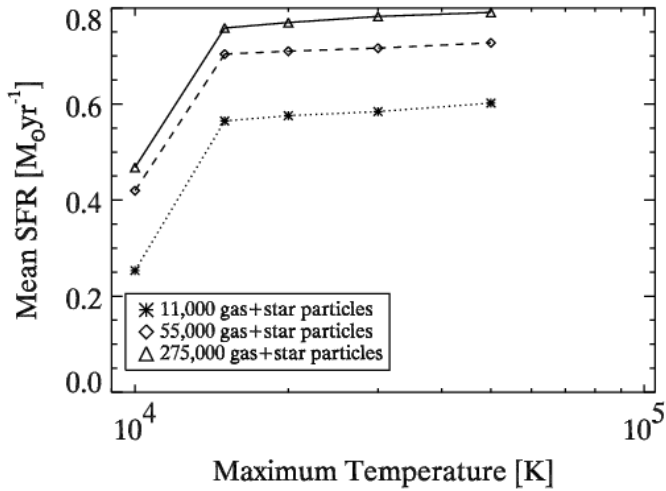}}}
\caption{The mean SFR as a function of $T_{\rm max}$ at low
resolution (asterisks), medium resolution (diamonds), and high
resolution (triangles). }  \label{fig:dTemp}
\end{center} \end{figure}

Figure \ref{fig:dTemp} shows the effect that varying $T_{\rm max}$ from
10,000-50,000 K has on the mean SFR.  As long as $T_{\rm max} \gta
15,000$ K the mean SFR is almost independent of $T_{\rm max}$.  We might expect so little effect since most gas particles that are near star forming temperatures have cooled all the
way to 10,000 K.  For example, only 18\% of the gas mass in the medium resolution simulation, and only 20\% of the gas mass at high resolution is warmer than 12,000 K after 1 Gyr of evolution using our fiducial recipe.  It is only those gas
particles that have been heated as the result of supernova feedback
that are excluded from forming stars by the $T_{\rm \rm max}$ criterion even
though they may remain in a dense environment.  We tried lowering the
threshold temperature all the way down to the mean gas particle
temperature, 10,000 K.  This produced galaxies with much burstier star
formation histories.  Gas particles would pile up just above the
temperature threshold, cool all at once, produce lots of star
formation, and then heat the gas up so that it could not form stars
until the next cooling episode started the cycle all over again.  This
experiment demonstrates that $T_{\rm \rm max}$ should stay above 12,000 K, but
its specific value does not critically affect star formation.  We
choose to use $T_{\rm max}=15,000$ K.  We note that the $T_{\rm max}$ criterion is
critical to our feedback prescription because it enables star
formation to be immediately suppressed by supernova feedback.

\subsubsection{$n_{\rm min}$: Minimum Density} 
\label{sec:rhomin} 
As $n_{\rm min}$ increases, fewer gas particles are eligible to form
stars, and hence fewer stars form.  Figure \ref{fig:
denmin} shows that a hundred-fold increase in the minimum
density causes an order of magnitude reduction in the star
formation rate.  As the gas density profile declines exponentially, most of the gas particles that are not eligible to form
stars reside in the outer parts of the disk.  Because the IMMW disk is
gravitationally stable, instabilities do not drive star
formation and thus density simply correlates with radius.
The minimum density that we choose, 0.1 cm$^{-3}$, confines star
formation the inner 20 kpc of our model galaxy, which corresponds to
surface densities of approximately 2$M_\odot$ pc$^{-2}$ and above and
matches the observed minimum density threshold for star formation in
nearby spiral galaxies \citep{MK01}.

\begin{figure} \begin{center} \resizebox{9cm}{!}{\resizebox{9cm}{!}{\includegraphics{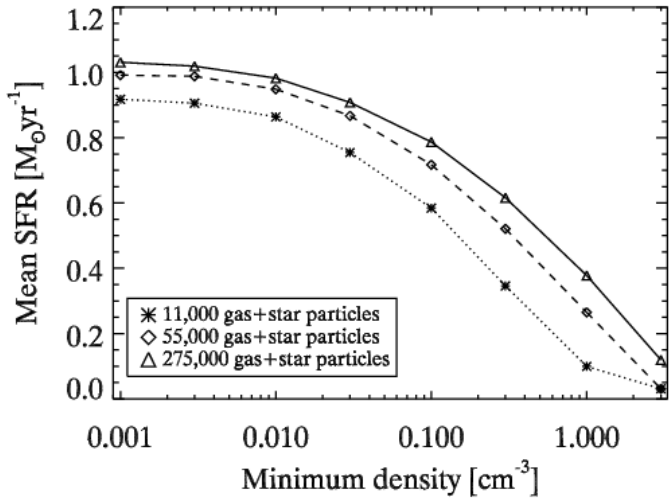}}}
\caption{The mean SFR as a function of minimum density
($\rho_{\rm min}$) at low resolution (asterisks), medium resolution
(diamonds), and high resolution (triangles). }  \label{fig:
denmin} \end{center} \end{figure}

\subsubsection{Jeans Criterion} The Jeans Criterion (\S
\ref{sec:SFCrit}) is a test of whether or not a gas cloud can provide
enough pressure support to prevent gravitational collapse and is
commonly used to test whether gas can form stars.  Our initial recipe
implemented it as a comparison between the sound crossing time,
$h_i/c_i$, where $h_i$ is the smoothing length and $c_i$ is the sound
speed, and the dynamical time, 1/$\sqrt{4\pi{G}\rho_i}$.

\begin{figure} \begin{center} \resizebox{9cm}{!}{\resizebox{9cm}{!}{\includegraphics{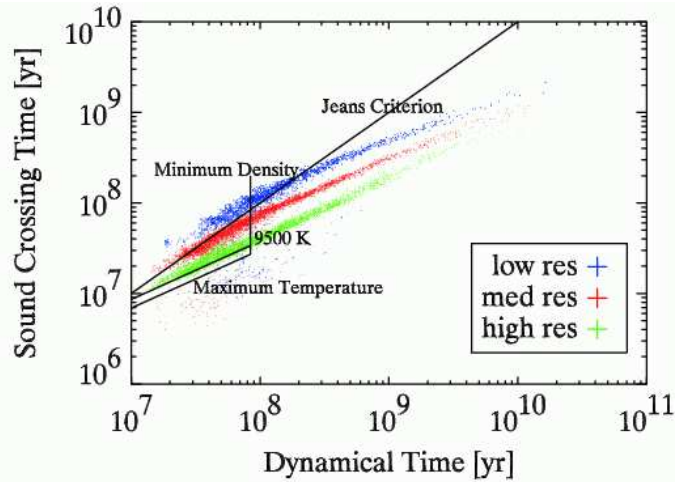}}}
\caption{The sound crossing time
across a smoothing length versus the local dynamical time
for 3 different simulation resolutions.
An increase in resolution,  which decreases the
smoothing lengths and hence the sound crossing times
takes gas particles from satisfying the Jeans Criterion in low
resolution simulations (blue points) to failing the criterion
in high resolution runs (green points) because the atomic cooling included in
these simulations only allows gas to cool to
10,000 K.  The line for particles at 9500 K uses typical smoothing
lengths of 360 pc at 0.1 ${\rm cm}^{-3}$ ($t_{\rm dyn}\sim80$ Myr) and 92 pc at
6 ${\rm cm}^{-3}$ ($t_{\rm dyn}=10$ Myr).  We also plot the maximum temperature
criterion for the high resolution simulation.
} \label{fig: jeansplot} \end{center} \end{figure}

Figure \ref{fig: jeansplot} shows this comparison for all the gas
particles in the simulation at three different resolutions.  What is
immediately apparent is that the sound crossing time decreases
significantly with increasing resolution,  which happens because a
region at given density is sampled by more particles in the higher
resolution simulations.  More resolution, without a corresponding drop
in temperature, results in a smaller smoothing length making the sound
crossing time shorter.  This decrease
in the sound crossing time becomes significant enough in the highest
resolution simulation to make gas particles unable to pass the Jeans
Criterion, at least using our formulation of it and
the atomic transition cooling function employed in
these simulations.  We also plot a line representing particles
at 9500 K in the high resolution simulation, as cold as they can get using
atomic cooling, using
typical smoothing lengths, which shows that typical particles have
little chance of passing the Jeans Criterion 
unless they cool a great deal adiabatically.  
Gas at the highest densities is sampled at the
highest resolution but because the gas does not realistically cool,
the sound crossing time becomes so short that t$_{\rm dyn}$ is still
greater than $h_i/c_i$, and the dense gas unrealistically fails the Jeans
Criterion. This might not be true if we included more realistic molecular
cooling.

\begin{figure} \begin{center} \resizebox{9cm}{!}{\resizebox{9cm}{!}{\includegraphics{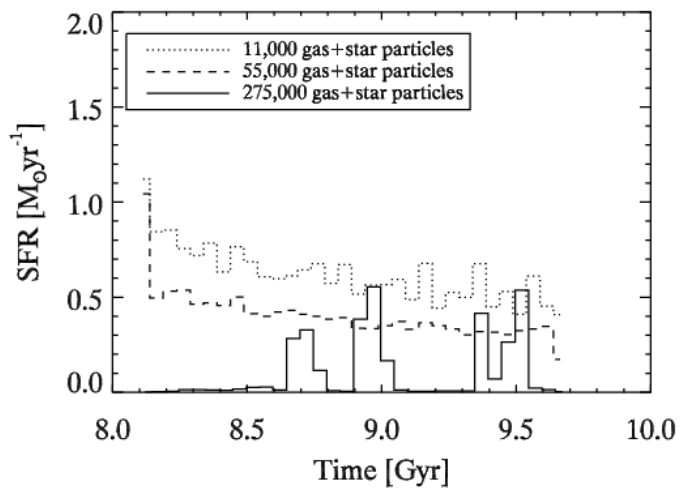}}}
\caption{The mean star formation rates at three
different resolutions when the Jeans criterion is enabled.
While the low ({dashed}) and medium ({dotted}) resolution
simulations form a similar number of stars, star formation is
nearly eliminated in the high
resolution ({solid}) simulation except for a couple of
bursts.}  \label{fig:jeanssfr} \end{center} \end{figure}

We plot the star formation histories obtained when we include the Jeans
Criterion in Figure \ref{fig:jeanssfr}. The consequences of this problem are
evident as few stars are able to form in the highest resolution simulation.
Thus, we reject using the Jeans Criterion
and use the simpler collapse conditions imposed by requiring that the gas 
have a minimum density and a maximum temperature.

Ideally the simulations would also include molecular cooling so that particles
would have more realistic temperatures, making the Jeans Criterion a
more appropriate test.  Figure \ref{fig: jeansplot} shows how the cooling curve
used in these simulations can only cool particles down to 9500 K at high
resolution.  Any further decrease in temperature is the result of adiabatic 
cooling.
Figure \ref{fig: jeansplot} also shows how the temperature maximum plays a
similar role to the Jeans criterion. However, the limit changes with resolution
in this plot because the sound crossing time decreases with decreasing smoothing
length while the sound speed of 15,000 K gas remains constant.  
For clarity, we only plot the high resolution maximum temperature limit.
In the current generation of simulations of forming galaxies, the models
only attempt to replicate the generic properties of star formation
that seem to be well represented by a Schmidt Law as observed by
\citet{Kenn98}.  A temperature maximum of $T_{\rm max}$=15,000 K ensures
that only reasonably cool gas particles form stars.  The stochastic
implementation of the star formation formula ensures the correct slope
and with $n_{\rm min}$= 0.1 cm$^{-3}$ we reproduce the correct
low density threshold, making a Jeans criterion unnecessary.

\subsubsection{Converging flow} Most of the particles that satisfy the
temperature and density criteria also satisfy the criterion that the
flow be converging.  Therefore, there is not a large difference in the
number of stars formed in our IMMW galaxy with the criterion turned on
or off.  The difference was minor enough that we decided to run all of
the simulations reported in this paper with the converging flow
criterion enabled since it might still be of benefit in nonequilibrium
situations that we will explore in future work.

\subsubsection{Summary of Criteria} The criteria that we settled on
are: \begin{itemize} \item The gas particle must be colder than
T$_{\rm max}$ = 15,000 K.  \item The gas particle must be denser than
$n_{\rm min}$ = 0.1 cm$^{-3}$.  \item The gas particle must be
overdense enough to be part of a virialised structure.  \item The
particle must be part of a converging flow, i.e.
$\nabla\cdot\textbf{v}<0$.  \end{itemize}

In addition we found that the Jeans Instability loses meaning at high
resolution as it is formulated in \citet{Katz92} and that star formation
is most sensitive to the $n_{\rm min}$ criterion.

\subsection{Effects of Parameters in the $\beta$ Model}
\label{subsec:beta} The criteria established in the above section are
appropriate for either of our feedback schemes.  In this section we
discuss how parameter choice affects star formation in the $\beta$
recipe (\S \ref{sec:betamodel}).  For all of these simulations, we
used our fiducial choices and changed only the specified parameter to
measure its effect on the star formation.

\subsubsection{$\beta$: Supernova Feedback Mass Factor} The $\beta$
parameter distinguishes our method from those used by \citet{Gerrit97}
and \citet{TC01}.  \citet{Gerrit97} turns off the radiative cooling
only for one gas particle while \citet{TC01} turn off the radiative
cooling for all 32 particles within the smoothing radius. The $\beta$
model gives us more flexibility.  For example, turning off the cooling
for all 32 neighbouring gas particles could produce a feedback that is
too strong and resolution dependent.  In a low resolution simulation,
the smoothing length is much longer than in a high resolution
simulation, so much more gas is affected and the feedback
is stronger.  However, this effect is mitigated to a certain extent
since the star particles are more massive, so that the feedback is the
result of more supernovae explosions.  Another possible problem is
that disabling the cooling for all 32 particles does not take into
account how much energy has been released in the supernova explosions.
Even if only one supernova explodes, it has the same effect as if
hundreds of supernova exploded.

\begin{figure} \begin{center} \resizebox{9cm}{!}{\resizebox{9cm}{!}{\includegraphics{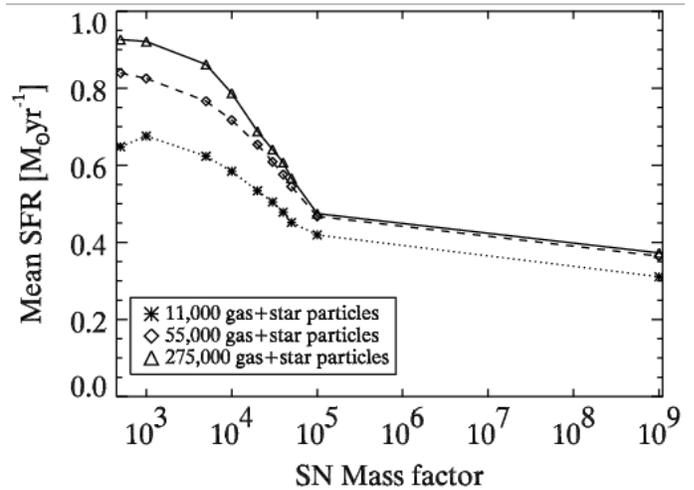}}}
\caption{The star formation rate as a function of mass factor $\beta$
at low resolution (asterisks), at medium resolution (diamonds), and at
high resolution (triangles).}  \label{fig:mfac} \end{center}
\end{figure}

As expected, Figure \ref{fig:mfac} shows that the more gas particles
that have their radiative cooling disabled, i.e. larger values of
$\beta$, the fewer stars form, because the gas particles are not able
to satisfy the maximum temperature criterion.  Once $\beta$
becomes large enough such that all the 32 neighbouring gas particles
that received supernova feedback energy have their cooling disabled,
increasing $\beta$ further has little additional effect.  The reduced
star formation at very low $\beta$ (the leftmost lowest resolution
point in Figure \ref{fig:mfac}) occurs because our analysis averages
star formation after the initial burst where much of the gas is
converted into stars leaving little gas to form stars at later times.

\begin{figure} 
\begin{center} 
\resizebox{9cm}{!}{\resizebox{9cm}{!}{\includegraphics{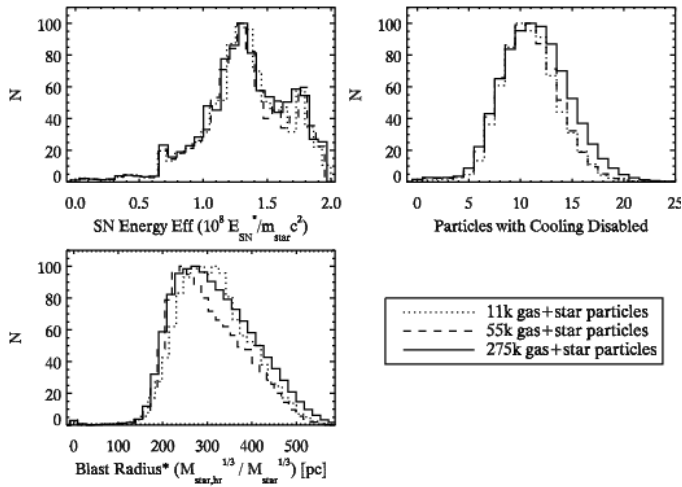}}}
\caption{Detailed characteristics of supernova feedback in the $\beta$
model for three different resolution simulations with the parameters set
at their fiducial values.  Each resolution has widely varying particle numbers, so every histogram has been normalized to a maximum of 100.  The statistics
were compiled over the first 200 Myr of the simulation.  The upper-left
panel plots the distribution of energies that each star particle releases from supernova explosions every time supernova feedback is calculated (1 Myr).  That quantity is normalised using star particle rest mass energies (${\rm E_{SN}^\star / m_\star c^2}$) to make comparisons between resolutions easier.  We note that the quantity ${\rm E_{SN}^\star}$, the energy the entire star particle releases into the ISM, is different than the quantity ${\rm E_{SN}}$ used elsewhere in this paper for the energy that individual supernovae release into the ISM.  Effectively, the normalised quantity represents the efficiency with which the matter in the star particles are converted into energy.  The upper-right panel plots the number of particles with their cooling disabled.  The relationship between gas particle mass and resolution means that the number of particles with cooling disabled roughly traces the gas mass with cooling disabled, which is similar for every resolution.
The lower-left panel plots the distribution of radii within which the
cooling is disabled renormalised to the high resolution simulation by
scaling by the mass in supernova to the one third power.}
\label{fig:snstats} \end{center} \end{figure}

Figure \ref{fig:snstats} shows the details of the $\beta$ feedback model
and can be compared with similar plots for the blastwave model shown
in Figure \ref{fig:bwstats}.  The plots show the results using the
fiducial recipe with $\beta$ = 10,000.  The critical panel is in the
upper, right-hand corner where we plot how many gas particles have
their cooling disabled for each star particle during a given star
formation event.  One can see that the number of particles affected is
not resolution dependent.  Star particles produce an amount of SN
energy proportional to their mass (as in the upper-left panel of Figure \ref{fig:bwstats}) and, therefore, one might expect the radii
within which we disable the cooling to be larger for lower resolution
simulations. However, the inter-particle separation also changes with resolution and thus the blastwave regions are proportionally the
same size as shown in the lower left hand panel.  Here we normalise
the radii to the high resolution simulation by scaling with the mass
in supernova to the one third power.  To recover the actual radii, in
the medium resolution simulation just multiply by 1.6 and in the low
resolution simulation by 2.7. But since the $\beta$ model does not
take into account how blastwaves expand differently in different
density environments,
the blastwave model provides a more physical representation
of supernova explosions as described in \S \ref{sec:bwrecipe}.

\subsubsection{$\tau_{\rm CSO}$: Cooling Shutoff Time } \label{sec:taucso}
\begin{figure} \begin{center}
\resizebox{9cm}{!}{\resizebox{9cm}{!}{\includegraphics{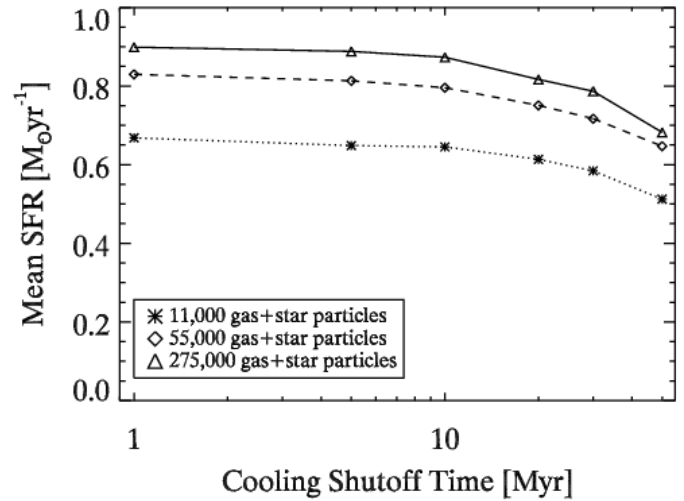}}} \caption{The mean SFR as a
function of the time that radiative cooling is disabled at low
resolution (asterisks), at medium resolution (diamonds), and at high
resolution (triangles).}  \label{fig:fbtime} \end{center} \end{figure}

We choose our fiducial $\tau_{\rm CSO}$=30 Myr, the time that we disable
the radiative cooling, based upon \citet{Gerrit97} and \citet{TC2000}.
Figure \ref{fig:fbtime} shows that there is not a significant increase
in the SFR as $\tau_{\rm CSO}$ decreases.  We prefer a short $\tau_{\rm CSO}$ as it more closely
matches the expected lifetime of SN blastwaves.

\subsubsection{$c^{\star}$: Star Formation Efficiency} $c^\star$
controls the efficiency with which stars form.  The $c^{\star}$
parameter can be thought of as either modifying the timescale for star
formation \citep{SH03} or as the fraction of gas that becomes
stars. The distribution of star formation timescales,
$t_{\rm form}$, for gas particles that pass the star formation criteria is close to a log normal distribution with a peak around 20 Myr and a tail out to 80 Myr.  The results are plotted for our fiducial
values at medium resolution and the times range from 20 to 80
Myr. Remember that $t_{\rm form}=t_{\rm dyn}= 1/\sqrt{4\pi G\rho}$ so the
times just trace the gas density.  Our fiducial value of
$c^{\star}$=0.1 either corresponds to star formation timescales of
$\sim 300$ Myr or implies that $\frac{1}{10}$ of a gas particle gets
converted to stars each star formation timescale, $\sim 30 Myr$.

\begin{figure} \begin{center} \resizebox{9cm}{!}{\resizebox{9cm}{!}{\includegraphics{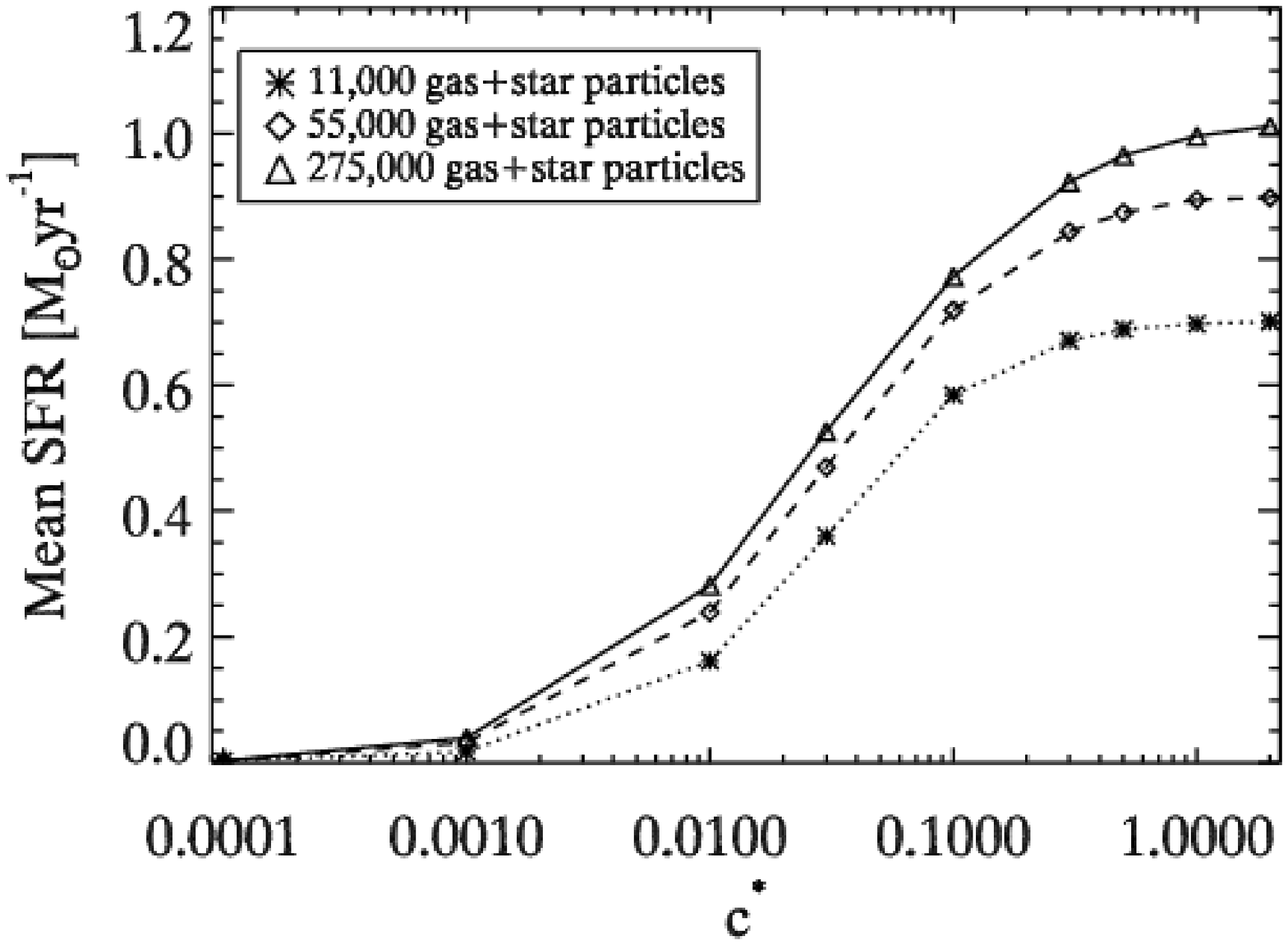}}}
\caption{The mean SFR as a function of c$^{\star}$ at low resolution
(asterisks), at medium resolution (diamonds), and at high resolution
(triangles).}  \label{fig:cstar} \end{center} \end{figure}

Figure \ref{fig:cstar} shows that $c^\star$ has the strongest effect
on star formation of any of the parameters that we have investigated
so far.  The value for $c^\star$ is thus tightly constrained using the
Schmidt Law \citep{Kenn98} defined in Equation \ref{eq:Kenn} and the
best fit for the $\beta$ feedback model is with $c^\star \sim 0.1$.

\subsubsection{$E_{\rm SN}$: Supernova Energy} During the late stages of
its life, the core of a massive star possesses $10^{53}$ ergs of
gravitational potential energy.  How much of that energy is converted
into thermal energy in the surrounding interstellar medium during a
Type II SN explosion remains an open question.  Most of the energy is
converted into neutrinos and a small fraction of the neutrino flux
interacts with enough matter to provide the canonical SN kinetic
energy value of $10^{51}$ ergs \citep{Colgate66}.  Integrating SN
light curves reveals that only $10^{49}$ ergs are radiated away in the
initial explosion \citep{Fil97}, but it is not clear that the rest of
the energy is transferred to the ISM.  In the current recipe, we have
chosen to leave the energy transfer efficiency as a parameter,
$E_{\rm SN}$, and use the \citet{Thornton98} estimate that 10$\%$ of the
supernova's kinetic energy is converted into thermal energy as our
fiducial value.  The rest of the energy is radiated away throughout
the lifetime of the supernova blastwave.

We distribute the supernova energy across all 32 nearest neighbour
particles using the smoothing kernel.  Since we are only disabling the
cooling in a limited set of those particles, some of the energy will
be quickly radiated away and have no effect on the simulation.
However, the majority of the energy goes to particles that do have
their radiative cooling disabled and thus the feedback does effect
star formation.

\begin{figure} \begin{center} \resizebox{9cm}{!}{\resizebox{9cm}{!}{\includegraphics{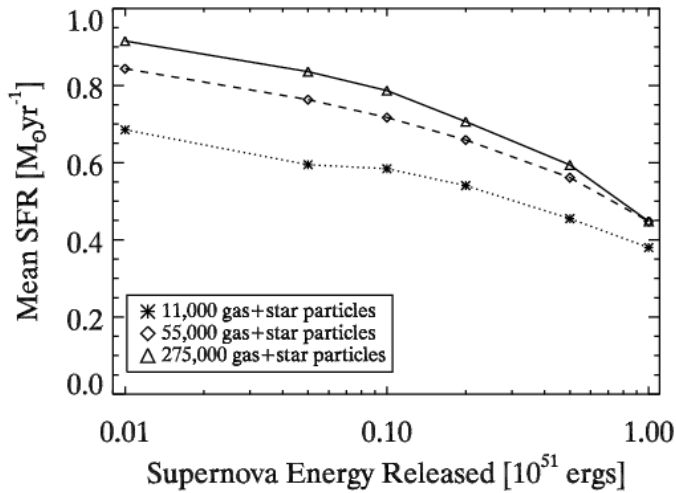}}}
\caption{The mean SFR as a function of $E_{\rm SN}$ using the $\beta$
feedback model at low resolution (asterisks), at medium resolution
(diamonds), and at high resolution (triangles).}  \label{fig:dESN}
\end{center} \end{figure}

Figure \ref{fig:dESN} shows that more stars form when less supernova
energy is returned to the ISM, as one would expect.  All of the
$E_{\rm SN}$ values came close to fitting the Schmidt Law, with $2\times10^{50} <
E_{\rm SN} < 6\times10^{50}$ doing the best.  $E_{\rm SN}$ does not have its largest
impact on star formation in our IMMW.  Its value is better constrained using other
observables (see \S\ref{sec:constraints}) or smaller galaxies.  The
reason that $E_{\rm SN}$ effects SFRs for the IMMW is that gas particles
get heated more when $E_{\rm SN}$ is increased and thus are less likely to fall
below the $T_{\rm max}$.  The gas may also expand more owing to the
additional pressure and be less likely to satisfy the $n_{\rm min}$
criterion.  We note that the star formation rate begins to converge to one mean SFR as
$E_{\rm SN}$ approaches $10^{51}$ ergs.

\subsection{Effects of Parameters on the Analytic Blastwave Model}
\label{subsec:blastwave}

The results of the previous section proved that turning off the
radiative cooling of gas particles is an effective means of
implementing supernova feedback in our IMMW.  The results also show that shorter
$\tau_{\rm CSO}$ and moderate $\beta$ values are in good agreement with
the analytic blastwave solutions presented in \citet{MO77}.  Using
these analytic expressions (detailed in \S \ref{sec:bwrecipe}) leaves
only two free parameters, $c^\star$ and $E_{\rm SN}$, governing both star
formation and feedback and is well motivated physically.

The results of two different versions of the blastwave model are
presented below.  First, supernova feedbacks, i.e. thermal energy,
metals, and mass, are distributed across the entire smoothing radius
as described in \S \ref{sec:largeSN}.  However, since energy that is
added to particles that can cool is lost almost immediately we propose
a second variant where we concentrate the feedback to only those
particles that have their cooling disabled as we describe in
\S\ref{sec:smallSN}.

\subsubsection{Smoothing over 32 Particles} \label{sec:largeSN} 
\begin{figure} \begin{center}
\resizebox{9cm}{!}{\resizebox{9cm}{!}{\includegraphics{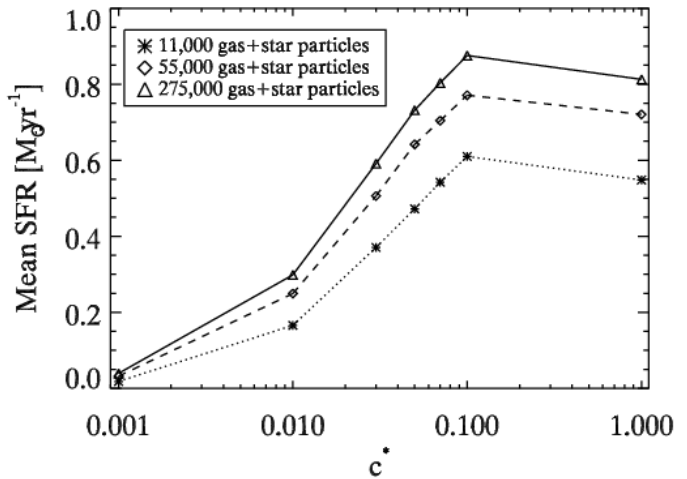}}} \caption{The mean SFR as a
function of c$^{\star}$ at three different resolutions using the
blastwave model and smoothing over 32 particles for simulations at low
resolution (asterisks), at medium resolution (diamonds), and at high
resolution (triangles).}  \label{fig:blastcstar} \end{center}
\end{figure}

The
star formation efficiency, $c^\star$, has a large effect on star
formation when we use the analytic blastwave model much like in the
$\beta$ model as we show in Figure \ref{fig:blastcstar}.  Most values
of $c^\star$ do not reproduce the observed Schmidt Law; only
$c^\star$=0.05 gives the \citet{Kenn98} normalisation at all three
resolutions.  Therefore, every simulation described hereafter uses
$c^\star$ = 0.05 as its fiducial value.

As discussed in \S\ref{sec:bwrecipe}, there are two possible shutoff
times in the blastwave method, $t_E$, the time that the snowplow phase
ends, and $t_{\rm max}$ the time that it takes the gas to cool down to the
ambient ISM temperature.  We find that in the IMMW simulations there
is little difference in the SFR between the two different choices with
each having a SFR of 0.8 $M_\odot$ yr$^{-1}$ at high resolution with
$c^\star$ = 0.05.

\begin{figure} \begin{center}
\resizebox{9cm}{!}{\resizebox{9cm}{!}{\includegraphics{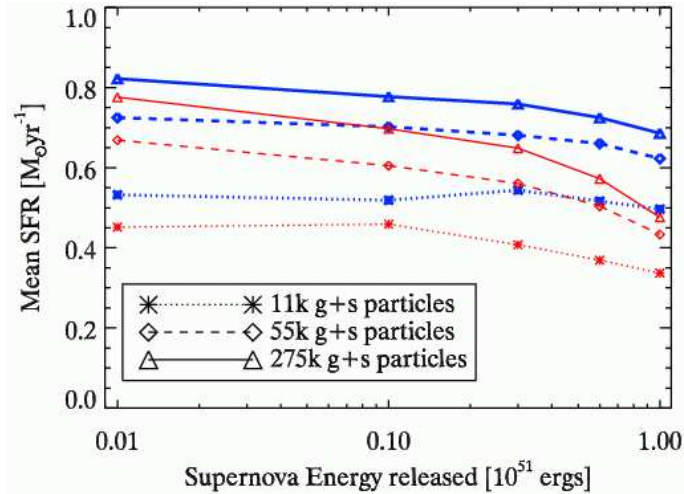}}} \caption{The mean SFR as a
function of supernova energy for the blastwave feedback model with
smoothing over 32 particles (thick, blue) and when the feedback is concentrated (thin, red) at low resolution (asterisks), at medium
resolution (diamonds), and at high resolution (triangles).}
\label{fig:blastdESN} \end{center} \end{figure}

The amount of supernova energy transferred to the ISM has little
effect on star formation as shown in Figure \ref{fig:blastdESN}.  All
of the values of $E_{\rm SN}$ using $c^\star$= 0.05 fit the observed
Schmidt Law well.  Supernova energy provides local feedback and this
maintains a steady SFR, but the amount of energy does not have a huge
impact on the amount of star formation.

\subsubsection{SN Ejecta Smoothed only over the Blast Radius}
\label{sec:smallSN} In our effort to inject the energy from supernovae
ejecta into the ISM both efficiently and more physically, we
concentrate the energy into just those particles inside of the blast
radius, i.e. those that will have their cooling disabled when SNII
explode.  We also concentrate the energy and metal deposition for SNIa
as a matter of convenience, but do not disable radiative cooling for
gas particles that are only within the SNIa blast radius.  When there
are no gas particles inside the blast radius, as often occurs when for
stellar wind feedback, we deposit all of the energy into the nearest
gas particle.

Concentrating the ejecta inside the blast radius has an
impact on star formation.  The average star formation rate drops to
0.7 $M_\odot$ yr$^{-1}$ at $c^\star$=0.05 and $E_{\rm SN} = 10^{50}$ ergs for the high resolution simulations
from the 0.8 $M_\odot$ yr$^{-1}$ shown previously.  While the star formation
decreases, the shape of the star formation history does not change.

Figure \ref{fig:blastdESN} shows how the mean SFR decreases more when the energy is concentrated within the
blast radius as opposed to spreading it over all 32 neighbouring
particles.  However, the variation in mean SFR does not lead to any deviation from the Schmidt Law, so it is not possible to constrain the value of $E_{\rm SN}$ from star
formation in the IMMW alone.

\subsection{Effects of Resolution} \label{sec:res} One trend that is
apparent from all the plots is that more stars form at higher
resolution.  With the fiducial recipe, there is a factor of two
difference in SFRs between the lowest resolution simulation with
11,000 particles and the highest resolution simulation with 275,000
particles.  There is a smaller difference between the medium and high
resolution simulations than there is between the low and medium
resolutions so perhaps the results are converging.  But a factor of
two remains a significant impediment to comparing simulations with
observations.

\begin{figure} \begin{center} \resizebox{9cm}{!}{\resizebox{9cm}{!}{\includegraphics{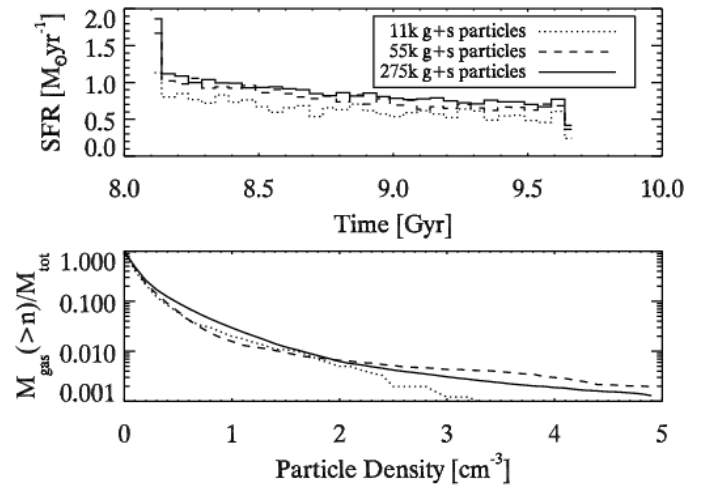}}}
\caption{The top panel shows how the SFR formation rate as a function of
time increases with increasing resolution for the fiducial parameter settings of the $\beta$ model at three different resolutions.  The bottom panel is a cumulative plot of gas mass fraction at various densities and shows why the SFR increases, as a greater fraction of gas is at star forming densities in the higher resolution simulations.}  \label{fig: resolution}
\end{center} \end{figure}

The difference in star
formation is caused in part by the higher densities that gas particles
reach in higher resolution simulations.  Denser gas forms more
stars because of Equation \ref{eq:starform}.  The bottom panel of Figure \ref{fig: resolution} shows that the most critical density range is between 0.5 and 1.5 $cm^{-3}$.  Most of the star forming eligible gas mass falls in this range and there is a significant difference in the fraction of gas with these densities.
In the high resolution simulation two times more gas has these densities than in the low resolution
simulation.  There are smaller fractions of gas at high density and Figure \ref{fig: resolution} shows that the different resolutions vary randomly in their high density gas mass content.  The top panel of Figure \ref{fig: resolution} shows that these variations have little bearing on how many stars form.

\begin{figure} \begin{center}
\resizebox{9cm}{!}{\resizebox{9cm}{!}{\includegraphics{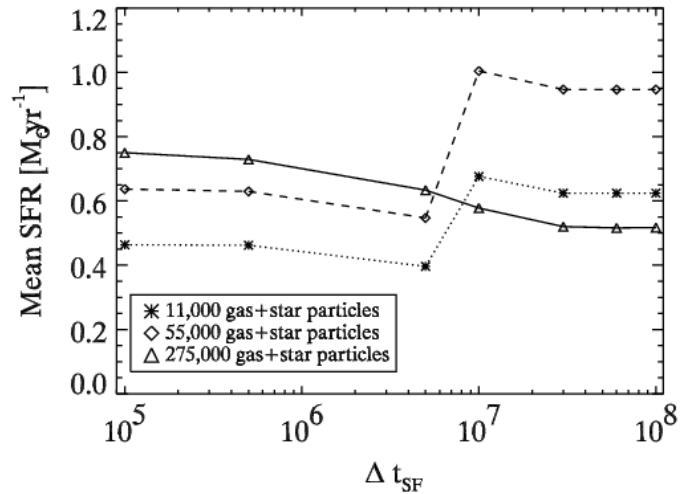}}} \caption{The mean SFR as a
function of the timestep used for star formation, $\Delta t_{\rm SF}$, for
the fiducial blastwave model at low resolution (asterisks), at medium
resolution (diamonds), and at high resolution (triangles).}
\label{fig: sftime} \end{center} \end{figure}

Time resolution can also affect the SFR; we calculate star formation
every ${\Delta}t_{\rm SF}$.  Our value of ${\Delta}t_{\rm SF}=1$ Myr comes
from considering the fact that O star lifetimes can be as short as 1 Myr.  Figure \ref{fig: sftime}, however, shows that the
actual value of ${\Delta}t_{\rm SF}$ is not crucial until either it starts
to approach the time over which we disable the cooling, calculated
using the analytic blastwave solution, which is typically around 10
Myr, or the length of a major system timestep, 15.6 Myr in these
simulations.  As star formation and feedback are not computationally
intensive, calculating it every Myr is practical and will not
adversely affect star formation.

\subsubsection{Feedback Behaviour} 
\begin{figure} \begin{center} \resizebox{9cm}{!}{\resizebox{9cm}{!}{\includegraphics{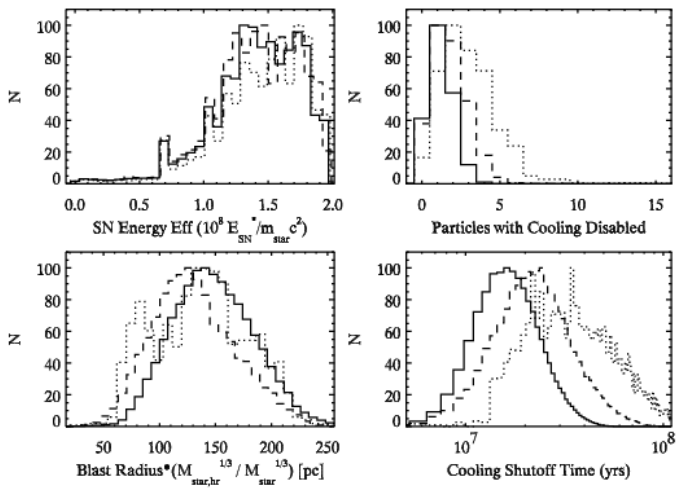}}}
\caption{The same as Figure
\ref{fig:snstats} for the blastwave feedback model where all the
feedback is concentrated within the blast radius.  There is one
additional panel in the lower right where we plot the distribution of
the cooling shutoff times.}  \label{fig:bwstats} \end{center}
\end{figure}

The feedback method also
contributes to the resolution sensitivity.  As in the $\beta$ method,
more particles have their cooling disabled when the blastwave recipe
is used at low resolution than at higher resolution as shown in Figure
\ref{fig:bwstats}.  The radial extent of the blastwave depends on the
energy of the explosion, so the bigger star particles in the low
resolution simulation generate a larger effect than the larger number
of star particles in the higher resolution simulations.  However, also
like in the $\beta$ method, the differences disappear if one rescales
by the mass in supernova to the one third power.  The length of time
that we disable cooling does depend on the resolution, i.e. the
cooling is disabled for longer periods at lower resolution.  This is
because supernova feedback events are larger but less frequent at
lower resolution.  At any one time there is about the same amount of
gas mass with its cooling disabled at all three resolutions.  It
should be noted that Figures \ref{fig:mfac}, \ref{fig:fbtime},
\ref{fig:dESN}, and \ref{fig:blastdESN} show that star formation
converges as the amount of feedback increases.  Unfortunately, these
values for the feedback are unphysical and produce fewer stars than
observed.

\section{Discussion} \label{sec:discussion} \subsection{Constraints}
\label{sec:constraints} The \citet{Kenn98} and \citet{MK01}
observations provide three constraints on how the star formation rate
relates to the gas surface density.  The first is that the logarithmic slope is 1.4$\pm0.2$ and constrains us to use the
stochastic formulation of star formation where SFR$\sim\rho/t_{\rm dyn}$.
The second is the normalisation of that relation.  Using the $\beta$
recipe, there are many different parameter combinations that can
produce the proper normalisation.  Using the blastwave method, only
$c^\star$ greatly influences the star formation rate and
$c^\star$=0.05 fits the \citet{Kenn98} observations the best.
Finally, there is the low density cutoff observed by \citet{MK01},
which we reproduce with our $n_{\rm min}$ criterion of $0.1 {\rm
cm}^{-3}$.

The observations of \citet{LocalSFR} indicate that stars have formed
at roughly a constant rate for some time in the Milky Way based on the
metallicity of G dwarfs.  In the isolated model Milky Way (IMMW) star
formation stays relatively constant with a slow decay when feedback is
enabled.  Figure 3 of \citet{LocalSFR} provides no conclusive evidence
of whether or not the star formation rate declines significantly.  The
major difference between our simulations and the Milky Way is that the
IMMW is isolated.  It has no external source of gas like the real
Milky Way, which may be in the form of gas rich mergers or cold gas
streaming in along filaments, and it is not gravitationally disturbed
by the presence of orbiting satellites.

\begin{figure} \begin{center}
\resizebox{9cm}{!}{\resizebox{9cm}{!}{\includegraphics{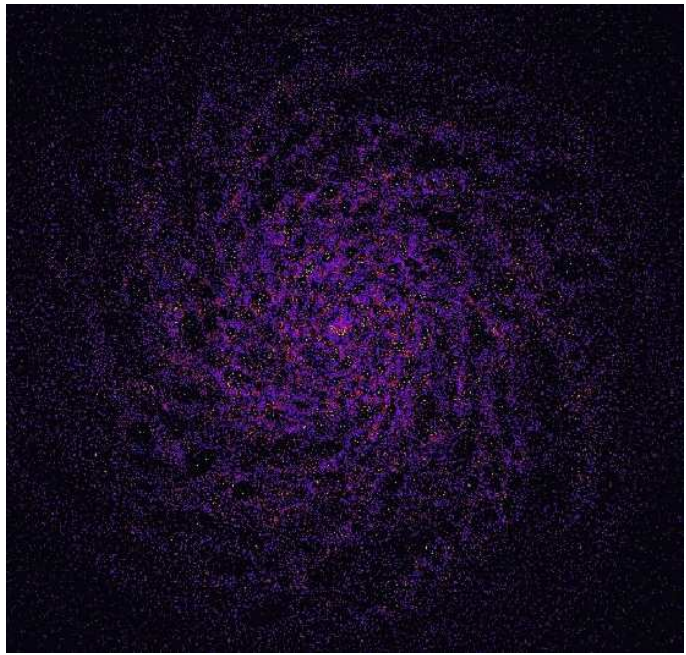}}} \caption{
An example of the effect that supernova
 feedback has on the structure of our IMMW.  The figure measures 24 kpc on a side.  Notice the hot, yellow
 particles that have pushed the colder, more purple particles more in the
 low density outer regions of the galaxy and less in the dense, inner
 regions.}  \label{fig:gaspic} \end{center} \end{figure}

Figure \ref{fig:gaspic} shows that our feedback method produces hot
gas particles that push the cold gas into dense filaments where the
gas particles are cool and dense enough to form stars.  The holes that
the SN blow open are smaller in the dense centre of our IMMW and grow
progressively larger as the SN explode in the less dense regions in
the outskirts of the galaxy.  Images of the Large Magellanic Cloud in
HI show similar features (see Figure 4 of
\citet{Jones99} or Figure 3 of \citet{Kim03} for a similar image of the Circinus Galaxy).  Inside
of the Milky Way disk, \citet{Hartmann02} observes that star formation
happens along filamentary structures.  Because of the jostling
provided by the adiabatic expansion of supernova heated gas, the site
of star formation is constantly shifting from one high density region
to the next high density region.  Our simulations are not run with
enough resolution to follow the fine turbulence and shockwaves that
lead to star formation, but do show structures that are reminiscent of
what is observed.

\begin{figure} \begin{center} \resizebox{9cm}{!}{\resizebox{9cm}{!}{\includegraphics{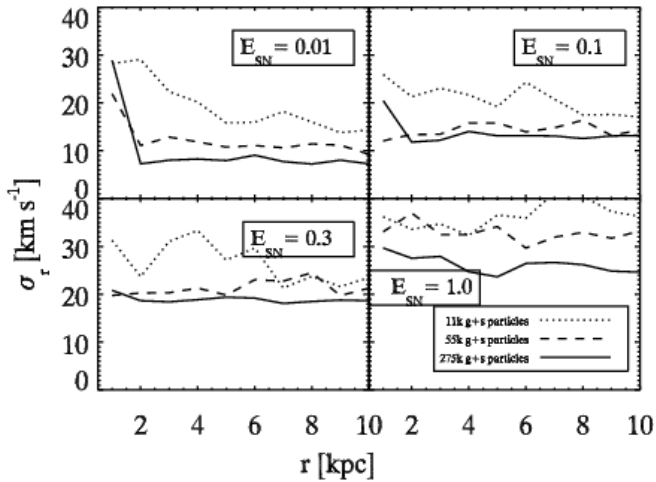}}}
\caption{The radial velocity dispersion vs. radius of the gas
particles in our concentrated blastwave model for four choices of the
supernova feedback energy, $E_{\rm SN}$ at the three resolutions.  We
calculate the dispersion within 10 evenly spaced, radial bins between
0 and 10 kpc.  The dispersion is most accurately determined for the
highest resolution simulation ({\emph solid line}).}  \label{fig:disp}
\end{center} \end{figure}

We cannot use star formation properties to constrain the one free
parameter in the blastwave feedback method, the amount of energy per
supernova, $E_{\rm SN}$, because in simulations
of the IMMW the SFR is not very sensitive to changes in $E_{\rm SN}$.  To
constrain this parameter we use the radial velocity dispersion of the
gas in the disk plane and the mass fraction of gas in the hot phase.
In Figure \ref{fig:disp} we plot the radial velocity dispersion of the
gas versus radius for the three resolutions using four different
values for $E_{\rm SN}$.  The dispersions are almost independent of radius
as observed in spiral galaxies. They are higher for the lowest
resolution simulation but similar for the two highest resolutions.  As
expected the dispersion increases with the energy of the feedback.
Typical vertical velocity dispersions of gas (under the assumption that random gas motions are isotropic) measured in quiescent
spiral galaxies are a little more than 10 km s$^{-1}$ \citep{DL90,DHH90}, so we
prefer $E_{\rm SN}\sim10^{50}$.

\begin{figure} \begin{center} \resizebox{9cm}{!}{\resizebox{9cm}{!}{\includegraphics{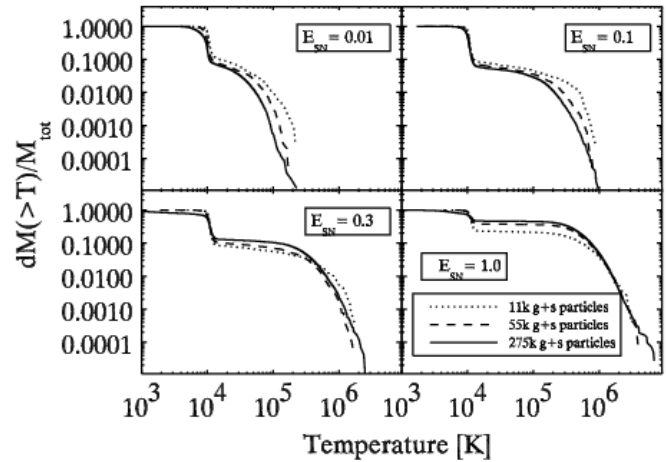}}}
\caption{The cumulative mass fraction of gas above a given temperature
for the blastwave method using four different supernova feedback
energies, $E_{\rm SN}$ at each of the three resolutions.}
\label{fig:temps} \end{center} \end{figure}

In Figure \ref{fig:temps} we plot the cumulative mass fraction of gas
above a given temperature.  We plot this at all three resolutions for
the same four choices of $E_{\rm SN}$ as in Figure \ref{fig:disp}.  The
temperature mass fractions are almost independent of resolution except
for the temperature of the hottest gas in the two cases with the
smallest supernova feedback. In these two plots the higher the
resolution the hotter the gas can become.  More importantly, the
fraction of gas in the cold star forming phase is very similar across
the three resolutions.  As expected, when the supernova feedback
energy becomes larger, the gas can become hotter and a smaller
fraction of gas remains in the cold phase.  Early soft x-ray
observations \citep{G74, B77} and observations of the O VI absorption
line \citep{JM74, York74} placed limits on the mass fraction of hot
gas in the Milky Way with a temperature greater than $5\times 10^5$ K.
Hot gas mass makes up $\sim0.01$ of the ISM \citep{vdH96}.  Since the
dispersion preferred value of $E_{\rm SN}=10^{50}$ agrees with these
observations, we fix the supernova energy at this value.

\subsection{Live Halo}

\begin{figure} \begin{center} \resizebox{9cm}{!}{\resizebox{9cm}{!}{\includegraphics{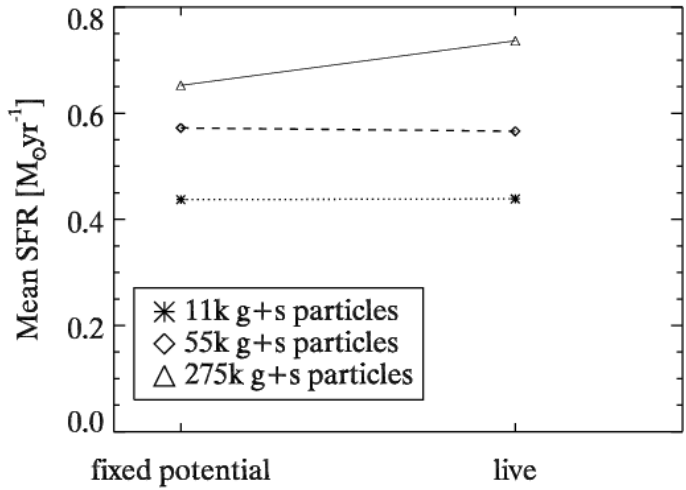}}}
\caption{The mean SFR for the final blastwave model both with and
without a live dark matter halo.}  \label{fig:live} \end{center}
\end{figure}

All the simulations reported so far have used a rigid analytic model
to represent the dark matter halo.  Live dark matter halos could
potentially introduce noise or allow secular instabilities to develop
that could change our SFRs.  To investigate this possibility we
resimulate the IMMW created using the \citet{Springel00} initial conditions, but for these simulations we do not remove the dark matter.  We use our fiducial concentrated blastwave feedback
method and all the fiducial parameter values for the live halo simulations.  We
evolve all three resolutions: for the high resolution simulation we
use one million dark matter particles to represent the dark halo and
use the same relative number of particles for the other two
resolutions.  As we show in Figure \ref{fig:live}, the addition of a
live halo does not significantly change the mean SFRs at any resolution.

\subsection{Summary of Chosen Parameters} \label{sec:sum} Based on the
constraints discussed in \S \ref{sec:constraints}, we determined the
parameter values in our star formation/supernova feedback recipe.  We
eliminated the Jeans Criterion used in \citet{Katz92} because it was
sensitive to variations in resolution.  Our parameter study using the
$\beta$ model converged towards the \citet{MO77} analytic blastwave
solution, so we chose to use the blastwave feedback method and
concentrated the feedback to those particles that have their cooling
disabled.  Employing this method means that there are only two free
parameters, $c^\star$ and $E_{\rm SN}$.  We constrained the star formation
efficiency, $c^\star$ to 0.05, which is close to the efficiency that
\citet{Lada+Lada} find in molecular clouds and star clusters.  The
mass of gas particles in the IMMW are close to the mass of molecular
clouds, so this value for $c^\star$ seems reasonable.  The amount of
supernova energy transferred to the ISM, $E_{\rm SN}$, is difficult to
constrain using SFRs in a massive system like the IMMW since the
galactic potential is so much deeper than the amount of energy that SN
can inject into the ISM.  The value of $E_{\rm SN}$ = $10^{50}$ ergs
comes from comparing gas velocity dispersions and temperature
distributions with observed values.

The blastwave model also requires that three criterion are met before
stars can form: $T_{\rm max}$, $n_{\rm min}$ and that the gas is in an
overdense virialised region.  Gas particles may only form stars when
their temperatures are below 15,000 K and is critical for making our
SN feedback mechanism effective.  Gas particles must also have a
density above $n_{\rm min}$=0.1 cm$^{-3}$.  The density criterion
limits star formation to the dense regions of galaxies, which
corresponds well to the density limits observed by \citet{MK01}.  The
\citet{Katz92} argument that gas particles be part of a converging
flow remains in our recipe because it does not have a major effect on
the number of stars that form and might be useful in nonequilibrium
situations.

\subsection{Comparison to Other Work} Several authors have recently
proposed alternatives to the star formation recipe described in this
paper.  Both \citet{Kravtsov03} and \citet{Li04} have run simulations
in which they do not impose a Schmidt Law for the star formation.
Instead, they obtain this result naturally.  While their studies shed
considerable light on how stars form in the environment of a galaxy,
our goal is to simply produce reasonable star formation rates in
simulations and thus relies on the observations of \citet{Kenn98}.

The \citet{Li04} recipe relies on high resolution simulations that at minimum satisfy the \citet{Bate+Burkert97} resolution criteria.  When simulations are run with an isothermal equation of state, the gas collapses because of instabilities into dense clumps where stars must form.  \citet{Li06} find that the collapse happens with the exact surface density dependence that \citet{Kenn98} observed.  \citet{Kravtsov03} notices a
steepening of the star formation rate with density in his simulations.
The steepening led him to use a constant star formation timescale
rather than a dynamical time that depends on density.
\citet{Kravtsov03} argues that this type of behaviour is the result of
a turbulent buildup of high density gas in such a way that the higher
the density of the gas, the more high density, star forming gas is
present.  \citet{Kravtsov03} only uses thermal energy from supernovae
as feedback, and hence it has little effect in his simulations.  His
need for altering the amount of star formation arises simply from how
the gas collapses, and his adaptive mesh code may be more effective at
resolving turbulence than our SPH code.  However, the idea of changing
the relationship between star formation and gas density could be
useful for creating a recipe with realistic star formation at very
high resolutions.

\citet{Tasker06} run a set of simulations similar to \citet{Li05} except that they include supernova feedback and the gas is simulated on a grid.  Even though their feedback implementation simply pours the feedback energy into the ISM, it proves effective at limiting star formation.  The effectiveness of the feedback may be the result of high resolution, effective shock capture, and a more realistic cool curve using the Enzo AMR grid program.  Even at the high resolutions, \citet{Tasker06} find it necessary to use a recipe like Equation \ref{eq:starform} that fixes the star formation dependence on density to a Schmidt law.

Recent observations also provide clues to a new physical recipe.
\citet{Blitz04} base a recipe on molecular gas observations.  The
recipe is much like \citet{E+E97} in that it identifies gas pressure
as the property most responsible for predicting SFRs.  Observations
also show interesting star formation behaviour in merging galaxies,
the place where the most vigorous star formation happens in the local
universe.  \citet{Barnes04} has pointed out that current star
formation recipes have a difficult time creating the starbursts
observed in the shocked regions of merging galaxies, since star
formation remains focused in the central regions of galaxies where the
densities are high enough to facilitate star formation.  It is yet to
be seen whether or not our star formation formulation suffers from
this problem.

\section{Conclusions} \label{sec:conclusion} Effective comparison of
simulations with observations requires that simulations include gas and
stars.  Computational limitations prevent simulations from
representing every atom or even every star in the universe.  However,
there are global properties of star formation \citep{Kenn98} that can
be reproduced in simulations.  The simulations presented here use a
scheme presented in \citet{Katz92} as a starting point.  The maximum
temperature, minimum density, converging flow, and stochastic
selection of star forming gas particles from the \citet{Katz92} model
were retained in our star formation recipe.  However, the Jeans
Criterion presented in that work showed a strong resolution dependence
in our simulations and was eliminated from our final star formation
method.

Simulations of an isolated model Milky Way (IMMW) that only included
this simple star formation recipe and pure thermal energy supernova
feedback had SFRs that were higher than those observed.  Therefore, we
implemented an effective version of supernova feedback that uses the
blastwave solution presented in \citet{MO77}.  Simulations cannot
resolve the blastwave shocks from supernovae.  Since star formation
occurs in dense gas regions, when the supernova energy is simply
distributed amongst nearby gas particles as thermal energy, it is
quickly radiated away and produces no effective feedback.  Thus, we
chose to disable the radiative cooling of gas particles in the
proximity of recently exploded supernovae so that the supernovae will
suppress star formation.  The \citet{TC01} simulations showed the
promise of this method, and we added the flexibility of how many gas
particles have their cooling disabled to eliminate the resolution
dependence of the \citet{TC01} method.

After an initial attempt using a recipe containing many parameters
(our $\beta$ model), we settled on the treatment of supernova
blastwaves presented in \citet{MO77} to determine how many particles
would have their cooling disabled.  This proved an effective means for
regulating star formation, and we were able to reproduce the Schmidt
Law observed by \citet{Kenn98} with $c^\star$=0.05.  Other than the
feedback effect necessary to make the IMMW's star formation constant,
the supernova feedback has little impact on star formation and we
expect it to have little effect in other massive systems.

Our final star formation recipe is made up of these parameter values:
\begin{itemize} \item $T_{\rm max}$ (maximum temperature) = 15,000 K \item
must be in a virialised region \item $n_{\rm min}$ (minimum density) =
0.1 cm$^{-3}$ \item must be in a converging flow, i.e.
$\nabla\cdot\textbf{v}<0$ \item $E_{\rm SN}$ (energy transfered from SN to
ISM) = $10^{50}$ ergs \item $c^\star$ (star formation
efficiency) =0.05 \item $R_E$ (SN blast radius) $=
10^{1.74}E_{\rm 51}^{0.32}n_0^{-0.16}\tilde{P}_{\rm 04}^{-0.20} pc$ \item
$t_{\rm max}$ (blast radius cooling time) $=
10^{6.85}E_{\rm 51}^{0.32}n_0^{0.34}\tilde{P}_{\rm 04}^{-0.70}$ {\rm yr}
\end{itemize}

Given the success of our star formation method in the isolated Milky
Way galaxy we hope it will continue in other situations.  In future
work we will investigate its performance when the galaxies have lower
mass and see whether or not our feedback prescription can blow
galactic winds like those observed.  We are testing it on much more
poorly resolved galaxies, like those that commonly arise in
cosmological simulations. Finally, we are applying this method to
cosmological simulations to see if we can reproduce the observed
shallow faint end slope of the galaxy luminosity function.

\section*{Acknowledgements} The anonymous referee made many detailed suggestions that greatly improved this paper.  We are grateful to Volker Springel for his
isolated galaxy model creation program.  The authors wish to thank
Julianne Dalcanton, Rob Thacker, Octavio Valenzuela, Yuexing Li, T.J. Cox, Joel Primack, Naomi
McClure-Griffths and Mordecai Mac Low for useful conversations.
Condor made managing all the simulations that we needed to run for
this paper much easier.  The Condor Software Program (Condor) was
developed by the Condor Team at the Computer Sciences Department of
the University of Wisconsin-Madison. All rights, title, and interest
in Condor are owned by the Condor Team.  The galaxy simulations were
run on machines funded by the Student Technology Fee of the University
of Washington.  Large simulations were run on the supercomputers
hosted at NCSA, ARSC, and University of Zurich.  Many thanks to
Joachim Stadel for allowing us to run some final, large simulations on
zBox2. The analysis of the simulations presented in this paper used
Salsa, funded by NASA ASIRP.  Funding for this project was provided by
NSF AST-0205969 and NASA NAGS-13308 \& NNG04GK68G.

\bibliographystyle{mn2e} \bibliography{SF}

 \end{document}